\documentclass[aps,pre, preprint]{revtex4-1}
\usepackage{amsthm}
\usepackage{subfigure}
\usepackage{graphicx}
\usepackage{epsfig}
\usepackage{epstopdf}
\usepackage{color}
\usepackage{caption}
\usepackage{amsmath}
\usepackage{amssymb}
\usepackage{hyperref}
\usepackage{soul}
\usepackage{wasysym}
\usepackage{ulem}
\newtheorem{theorem}{Theorem}[section]
\theoremstyle{definition}
\newtheorem{defn}[theorem]{Definition}

\newcommand\solidrule[1][1cm]{\rule[0.5ex]{#1}{.4pt}}

\begin{document}
\title{Onset of fingering instability in a finite slice of adsorbed solute}

\author{Tapan Kumar Hota, Satyajit Pramanik, Manoranjan Mishra}
\affiliation{Department of Mathematics, Indian Institute of Technology Ropar, Nangal Road, 140001 Rupnagar, India}

\begin{abstract}
The effect of a linear adsorption isotherm on the onset of fingering instability in a miscible displacement in the application of liquid chromatography, pollutant contamination in aquifers etc. is investigated. Such fingering instability on the solute dynamics arise due to the miscible viscus fingering (VF) between the displacing fluid and sample solvent. We use a Fourier pseudo-spectral method to solve the initial value problem appeared in the linear stability analysis. The present linear stability analysis is of generic type and it captures the early time diffusion dominated region which was never expressible through the quasi-steady state analysis (QSSA). In addition, it measures the onset of instability more accurately than the QSSA methods. It is shown that the onset time depends non-monotonically on the retention parameter of the solute adsorption. This qualitative influence of the retention parameter on the onset of instability resemblances with the results obtained from direct numerical simulations of the nonlinear equations. Moreover, the present linear stability method helps for an appropriate characterisation of the linear and the nonlinear regimes of miscible VF instability and also can be useful for the fluid flow problems with the unsteady base-state.
\end{abstract}

\maketitle

\section{Introduction}\label{sec:introduction}
The displacement process in porous media has enormous importance in the field of fluid dynamics. It features viscous fingering (VF) instability when a more viscous fluid is displaced by a less viscous one \cite{Homsy1987}. A confined geometry of Hele-Shaw cell \cite{HeleShaw1898} is generally used to experimentally observe VF in a homogeneous porous medium. Homsy presented an insightful review on VF instability in both miscible and immiscible fluids \cite{Homsy1987}. Due to its application in many industrial and environmental processes, such as the secondary oil recovery from porous rocks \cite{Engelberts1951}, pollutant contamination in underground aquifers \cite{Brian2008}, etc., this problem has drawn attention of multidisciplinary researchers for many decades. The objectives of these researchers are multiple, such as, understanding morphological instability leading to interfacial pattern formation \cite{Hill1952,  Slobod1963, Perkins1965, Tan1988}, to present a suitable stability analysis of hydrodynamic instabilities with unsteady base-state \cite{Tan1986, Ben2002, Pritchard2009, Kim2012, Pramanik2013, Gandhi2014}, and many more. VF is also observed in liquid chromatography, a flow based separation method in which a given fluid (called the displacing fluid) displaces a miscible sample consisting of a solvent and a mixture of dissolved solutes (called analyte) \cite{Guiochon2008}. 

Mathematical challenges of performing linear stability analysis (LSA) for this type of hydrodynamic instability is that the base state is not translated unchanged along the flow direction as time elapses. Diffusion relaxes the interface between the two fluids, resulting this as an unsteady base-state problem. The challenge is to capture the linearly unstable modes and their onset of instability by suitably incorporating the time evolution of the base-state. In order to overcome this difficulty Tan and Homsy \cite{Tan1986} used a quasi-steady state approximation, which assumes that the growth rate of the disturbances is faster than the rate of change of the base state. Following Tan and Homsy \cite{Tan1986}, Rousseaux {\it et al.} \cite{Rousseaux2007} performed a linear stability analysis to capture the onset of instability in liquid chromatography column using QSSA. Although QSSA method successfully measures the growth rate of the perturbations with a certain degree of accuracy, all the perturbations are found to be unconditionally unstable for arbitrary small time $t > 0$. Therefore, QSSA method fails to fulfil the most important goal of stability analysis, which is to appropriately measure the onset of instability. Growth rate measured from the QSSA method is valid for large time only. Thus, one needs to look for an alternative method for linear stability analysis to predict the onset of instability accurately. In this direction Ben {\it et al.} \cite{Ben2002} performed an LSA using spectral analysis and showed an unconditional stability of the system at early times. Recently, Kim \cite{Kim2012}, Pramanik and Mishra \cite{Pramanik2013} applied QSSA method in a self-similar domain to calculate the onset of VF instability in miscible slices. These authors also found unconditional stability at early times, which is in accordance with the spectral analysis \cite{Ben2002}. Contrary to classical single interface VF \cite{Tan1986}, the base state concentration profile for the displacement of a finite slice is not self-similar. Apart from the QSSA method and spectral analysis, another way to perform an LSA of problems with an unsteady base state is to solve the linearized equations as an initial value problem (IVP) \cite{Tan1986}. The obstacle in this method is that the representative initial condition is not known a priori and often the `white noise' or random perturbation is used as the initial condition. The main difficulty arises with such an initial condition is that it introduces perturbation to the entire system. One of the aims of this paper is to present a generic linear stability analysis for flow with unsteady base-state such that it captures all the physics accurately and also is computationally more efficient. 

Adsorption of solute on the porous matrix influences the separation in liquid chromatography and pollutant contamination in aquifers \cite{Brian2008, Keunchkarian2006}. A theoretical model incorporating linear adsorption of the species ruling the dynamic viscosity of the solution in a miscible displacement was analyzed by Mishra {\it et al.} \cite{Mishra2007}. These authors presented an LSA using QSSA method and also direct numerical simulations (DNS) of the nonlinear equations. Their study revealed that in the presence of a retention governed by $\kappa'$, the effect of solute is similar to that of an unretained solute with the dynamic viscosity reduced by  a factor of $1+\kappa'$. To incorporate the dynamics of the carrier fluid and of both the solute and the solvent, Mishra {\it et al.} \cite{Mishra2009} proposed a theoretical model with viscosity being ruled by sample concentration and solute being a  passive scalar adsorbed linearly. They performed DNS using a Fourier pseudo-spectral method to study the effect of VF on spatio-temporal distribution of the retained solute and the influence of the retention parameter on the fingering of solute concentration. Recently, Mishra {\it et al.} \cite{Mishra2013}, Rana {\it et al.} \cite{Rana2014a, Rana2014b} numerically studied the influence of solvent modulated adsorption on the propagation dynamics of the solute in the absence and presence of viscosity contrast, respectively. 

In this context, we perform a linear stability analysis using a Fourier pseudo-spectral method to have a more comprehensive understanding about the influence of solute adsorption on the onset of instability. For simplicity, we consider the retention to be independent of the solvent concentration. Unlike modal analysis, the present LSA calculates separate growth rates for the perturbations associated with each of the physical quantities. It also captures the diffusion dominated region at the early time and  distinguishes the linear and nonlinear regimes, which were never achieved through the existing LSA methods for problem with the unsteady base-state. We show that the linear unstable modes for the sample solvent remain unaffected by the adsorption of the solute on the porous matrix. It is further identified that the onset of fingering instability of the adsorbed solute has a non-monotonic dependence on the retention parameter. These results are consistent with the direct numerical simulations of Mishra {\it et al.} \cite{Mishra2009}. 

The paper is organized as follows. We present the mathematical model of the problem in Sec. \ref{sec:MM}, followed by the stability analysis and numerical method of solution of the problem in Sec. \ref{sec:LSA_DNS}. Secs. \ref{sec:NoAdsorp} and \ref{sec:Adsorption} discuss the obtained results without and with the adsorption of the solute concentration on porous matrix, respectively. In these sections, the LSA and DNS results are compared, followed by concluding remarks in Sec. \ref{sec:conclusion}. 

\begin{figure}[h!]
\centering
\hspace*{-0.7cm}\includegraphics[width=3in, keepaspectratio=true, angle=270]
{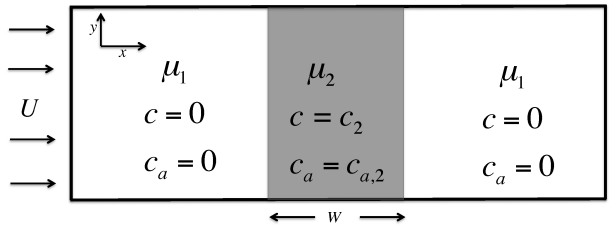} \hspace*{0cm}
\setlength{\abovecaptionskip}{-3cm}
\caption{Schematic of the flow configuration for three component model with coordinate axes.}\label{figure_1}
\end{figure}

\section{Mathematical Model}\label{sec:MM}
Consider a uniform rectilinear displacement of a sample solvent of width $W$ and viscosity $\mu_2$ injected at initial time $t = 0$ in a two-dimensional, homogeneous, horizontal porous medium (or a Hele-Shaw cell) by a carrier fluid or eluent of viscosity $\mu_1 ~(< \mu_2)$ (see Fig. \ref{figure_1}). The permeability of the porous medium is assumed to be constant $K$, which is equivalent to $b^2/12$ for a Hele-Shaw cell consisting of two parallel plates separated by a small gap $b \ll H$, where $H$ is the width of the Hele-Shaw plates. The sample consists of a solute or analyte of concentration $c_a = c_{a,2}$ dissolved in a solvent of concentration $c = c_2$. The sample solvent is different from the carrier fluid in which the sample solvent concentration $c = 0$ and the solute concentration is $c_a = 0$. The eluent is injected with a uniform velocity $U$ along the $x$ direction from left to right as shown in Fig. \ref{figure_1}. 

Assuming that the fluids are neutrally buoyant, incompressible and the dispersion is isotropic we can describe the above-mentioned three component model using the non-dimensional equations in a Lagrangian frame of reference moving with the velocity $U$ \cite{Mishra2009}, 
\begin{eqnarray} 
\label{cont_eqn}
& & \nabla \cdot \vec{u} = 0, \\ 
\label{darcy_eqn}
& &\nabla p = -\mu(c) (\vec{u} + e_x),\\
\label{convec_diffuse}
& & \frac{\partial c} {\partial t} + \vec{u} \cdot \nabla c = \nabla^2 c,\\
\label{solute}
& & (1+\kappa')\frac{\partial c_{a,m}} {\partial t} + (\vec{u} - \kappa' e_x) \cdot \nabla c_{a,m} = \nabla^2 c_{a,m}, 
\end{eqnarray}
where $p$ is the dynamics pressure, $ \vec{u} = (u,v)$ is the two dimensional gap-averaged velocity in the Lagrangian frame of reference, $\mu(c)$ is the dynamics viscosity of the fluids, and $e_x$ is the unit vector in the $x$ direction. Further, $c$ and $c_{a,m}$ correspond to the solvent concentration and the mobile phase solute concentration, respectively. For the non-dimensionalization we use $U, D/U,$ and $D/U^2$, respectively, as the characteristic velocity, length and time. Here $D$ corresponds to the isotropic dispersion tensor. The reference concentration for the solvent and the solute concentration are taken as $c_2$ and $c_{a,2}/(1 + \kappa')$, respectively. The log-mobility ratio $R = \ln(\mu_2/\mu_1)$, retention parameter $\kappa'$ and the dimensionless sample width $w$ are the three dimensionless parameters of the problem. The viscosity and concentration are related by an Arrhenius type relationship \cite{Homsy1987}, i.e., $\mu(c) = e^{Rc}$. The initial and boundary conditions associated with Eqs. \eqref{cont_eqn}-\eqref{solute} are: 
\begin{eqnarray}
\label{IC_1}
& & \vec{u}(x,y,t = 0) = (0,0), \\
& & c(x,y,t=0) = c_{a,m}(x,y,t=0)=\begin{cases}
1, & 0 \leq x\leq w\\
0, & ~ \mbox{otherwise}, 
\end{cases}
\end{eqnarray} 
and 
\begin{eqnarray}
\label{bc_1}
& & \vec{u} = (0,0),~\left(\frac{\partial c}{\partial x}, \frac{\partial c_{a,m}}{\partial x}\right) \rightarrow  (0,0), ~ \text{as} ~ |x| \rightarrow \infty,\\
\label{bc_2}
& &\frac{\partial c}{\partial y} = \frac{\partial c_{a,m}}{\partial y} = \frac{\partial v}{\partial y} = 0, ~\forall x,  
\end{eqnarray}
respectively, where $w$ corresponds to the dimensionless width of the sample. The velocity boundary condition in Eq. \eqref{bc_2} corresponds to constant pressure at the spanwise boundaries, where the streamwise velocity component $u$ takes arbitrary value \cite{Nield1992}. 

\subsection{Stream function formulation}
For a two dimensional flow, the  conservation of mass is satisfied by introducing the stream function, $\psi(x,y,t)$, such that the velocity components are given by $u = \psi_y,
\; v = -\psi_x$. Now taking the curl of the non-dimensional Darcy's equation (Eq. \eqref{darcy_eqn}) the pressure is eliminated, and we obtain
\begin{eqnarray}
\label{eq:SV1}
& & \nabla^2\psi = -R\left(\nabla\psi + \frac{\partial c}{\partial y} e_y\right) \cdot\nabla c , \\
\label{eq:SV2}
& & \frac{\partial c}{\partial t} + \left( \frac{\partial \psi}{\partial y}, \; -\frac{\partial \psi}{\partial x}\right)\cdot \nabla c  = \nabla^2 c, \\
\label{eq:SV3}
& & (1+\kappa')\frac{\partial c_{a,m}}{\partial t}  +\left( \frac{\partial \psi}{\partial y} - \kappa', -\frac{\partial \psi}{\partial x} \right) \cdot \nabla c_{a,m} \nonumber \\
& & ~~~~~~~~~~~~~~~~~~~~~~~~~~~~~~~~~~ = \nabla^2 c_{a,m}, 
\end{eqnarray} 
where $e_y$ is the unit vector in the $y$ direction. The initial condition and the longitudinal and transverse boundary conditions corresponding to the velocity are expressed in terms of the stream function as, 
\begin{eqnarray}
& & \frac{\partial\psi}{\partial y} = \frac{\partial\psi}{\partial x} = 0, ~~~ \text{at}~ t = 0, \\
& & \left(\frac{\partial\psi}{\partial y}, \frac{\partial\psi}{\partial x}\right) = (0,0), ~~~ \text{as}~ |x| \to \infty, \\
& & \frac{\partial^2\psi}{\partial y\partial x} = 0 ,  ~ \forall x ,
\end{eqnarray}
respectively. 
 
\section{Stability analysis and numerical solutions}\label{sec:LSA_DNS}
In this section, the unsteady base state of the model is presented, followed by the derivation of the linearized perturbation equations. Further, the numerical solutions of the linear stability problem as well as the direct numerical simulations of the fully nonlinear problem using a highly accurate Fourier pseudo-spectral method are presented. The growth rate and the hence onset of instability is obtained by projecting the governing equations as an initial value problem. 

\subsection{Base state}\label{subset:basestate}
For base-state flow we assume $\vec{u}^b = (0,0)$ which implies $\psi^b$ to be constant. We also assume that the base-state concentration is homogeneous in the $y$-direction, i.e., $c^b = c^b(x,t)$ and $c_{a,m}^b = c_{a,m}^b(x,t)$. Using Fourier transform and Eq. \eqref{IC_1}, the base-state flow can be written in terms of decaying error function solution of step-like initial concentration profiles for the solvent and solute, 
\begin{eqnarray}
\label{concen_base_state}
& & c^b(x,t) = \frac{1}{2}\left[\text{erf} \left(\frac{x}{2\sqrt{t}}\right) - \text{erf} \left(\frac{x - w}{2\sqrt{t}}\right)\right], \\
\label{solute_base_state}
& & c_{a,m}^b(x,t) = \frac{1}{2}\left[\text{erf} \left(\frac{x + \kappa' t/(1 + \kappa')}{2 \sqrt{t/(1 + \kappa')}}\right)\right. \nonumber \\
& & ~~~~~~~~~~~~~~~~~~~ \left. - \text{erf} \left(\frac{x + \kappa' t/(1 + \kappa') - w}{2 \sqrt{t/(1 + \kappa')}}\right)\right], 
\end{eqnarray}
respectively. It is clear from Eqs. \eqref{concen_base_state} and \eqref{solute_base_state} that the basic state of the stability problem is a diffusing front which is not stationary. To enable modal stability theory, the disturbance is often decomposed as Fourier modes in the transverse direction. In this process each Fourier mode is also assumed to be decoupled which contradicts the empirical observation that the concentration disturbances are localized in the downstream direction within the diffusive layer. We know that the concentration gradient at the front scales as $\sqrt{t}$ for one that begins as a step profile. Hence, unless the front has been allowed to diffuse initially before displacement begins, the rate of change of the base-state can be arbitrarily large for $t \ll 1$. In a self-similar $(\xi,y,t)$ coordinate, the base state equation for sample solvent $c^b$ becomes 
\begin{eqnarray}
\label{concen_base_state_SS}
& & c^b(\xi,t) = \frac{1}{2}\left[\text{erf} \left(\frac{\xi}{2}\right) - \text{erf} \left(\frac{\xi}{2} - \frac{w}{2\sqrt{t}}\right)\right],
\end{eqnarray}
where $\xi= x/\sqrt{t}$ is the similarity variable. In contrast to the single interface, the base state in the present case is not self-similar i.e., $c^b$ is dependent on both $\xi$ and $t$ and the transient effect of the base state can be no more negligible. Thus to avoid such barriers in modal linear stability analysis, QSSA is not invoked in this paper. 

\subsection{Linearized perturbed equations}\label{sec:LSA}
The principle of an LSA is to observe whether infinitesimal perturbations introduced to the equilibrium state (Eqs. \eqref{concen_base_state} and \eqref{solute_base_state}) amplify or decay in  time. For this, perturbations to the base-state flow, such that $c(x,y,t) = c^b(x,t) + \theta c'(x,y,t)$, $c_{a,m}(x,y,t) = c_{a,m}^b(x,t) + \theta  c_{a,m}'(x,y,t)$ and $\psi(x,y,t) = \psi^b + \theta \psi'(x,y,t)$, etc. are introduced. We already mentioned in Sec. \ref{subset:basestate} that $\psi^b$ is constant, which can be assumed to be equal to zero without loss of generality. On substituting these expressions Eqs. \eqref{eq:SV1}-\eqref{eq:SV3} can be written in terms of perturbation quantities as, 
\begin{eqnarray}
\label{eq:NLP1}
& & \theta \left[\nabla^2\psi' + R\frac{\partial c^b}{\partial x}\frac{\partial \psi'}{\partial x} + \frac{\partial c'}{\partial y} \nonumber \right]\\
& & ~~~~~~~~~~~~~~~~~ + \theta^2 R \left[\frac{\partial c'}{\partial x}\frac{\partial \psi'}{\partial x} + \frac{\partial c'}{\partial y}\frac{\partial \psi'}{\partial y}\right]= 0,\\
\label{eq:NLP2}
& & \theta \left[\frac{\partial c'}{\partial t} + \frac{\partial c^b}{\partial x}\frac{\partial \psi'}{\partial y} - \nabla^2 c'\right] \nonumber \\
& & ~~~~~~~~~~~~~~~~~ + \theta^2\left[\frac{\partial c'}{\partial x}\frac{\partial \psi'}{\partial y} - \frac{\partial c'}{\partial y}\frac{\partial \psi'}{\partial x}\right] = 0, \\
\label{eq:NLP3}
& & \theta\left[(1 + \kappa')\frac{\partial c'_{a,m}}{\partial t} + \frac{\partial c_{a,m}^b}{\partial x}\frac{\partial \psi'}{\partial y} - \kappa'\frac{\partial c'_{a,m}}{\partial x} -  \nabla^2 c'_{a,m} \right]\nonumber \\
& & ~~~~~~~~~~~~~~~~~ + \theta^2\left[\frac{\partial c'_{a,m}}{\partial x}\frac{\partial \psi'}{\partial y} - \frac{\partial c'_{a,m}}{\partial y}\frac{\partial \psi'}{\partial x}\right]  = 0.
\end{eqnarray}
In an LSA the perturbations are required to be infinitesimally small, which can be obtained by assuming $|\theta| \ll 1$. Neglecting the terms containing $\theta^2$ and equating the coefficients of $\theta$ equal to zero, we obtain the corresponding linearized equations in terms of the perturbation quantities, 
\begin{eqnarray}
\label{eq:LP1}
& & \nabla^2\psi' + R\left(\frac{\partial c^b}{\partial x} \frac{\partial \psi'}{\partial x} + \frac{\partial c'}{\partial y}\right) = 0,\\
\label{eq:LP2}
& & \frac{\partial c'}{\partial t} + \frac{\partial c^b}{\partial x} \frac{\partial \psi'}{\partial y} - \nabla^2 c' = 0, \\
\label{eq:LP3}
& & \frac{\partial c'_{a,m}}{\partial t} + \delta  \frac{\partial c_{a,m}^b}{\partial x} \frac{\partial \psi'}{\partial y} + \lambda \frac{\partial c'_{a,m}}{\partial x}  - \delta\nabla^2 c'_{a,m} = 0,
\end{eqnarray} 
where $\delta = 1/(1 + \kappa')$ and $\lambda = -\kappa'/(1 + \kappa')$. The boundary conditions associated with the equations in perturbation quantities are,
\begin{eqnarray}
\label{eq:perturb_BC1}
& & (c', c'_{a,m}, \psi')(x,y,t) = (0,0,0), ~ \text{at}~ x = 0~ \text{and}~ L_x, \\
\label{eq:perturb_BC2}
& & (c', c'_{a,m}, \psi')(x,y,t) = (0,0,0), ~ \text{at}~ y = 0~ \text{and}~ L_y, 
\end{eqnarray}
where $L_x$ and $L_y$ represent the dimensionless length and width of the computational domain, respectively. In order to understand the instability phenomenon we use different initial conditions, and it will be described in Sec. \ref{subsec:AGRP}. 

\subsection{Initial value problem and Fourier pseudo-spectral method}\label{subsec:IVP_FSM}
The traditional approach for studying the LSA for unsteady base state is by frozen-time approach which is known as QSSA. In the present case, QSSA approach can not predict the growth rate of individual flow variables (see Appendix \ref{subset:QSSA}), and hence it can not meet our principal aim i.e., analysing the evolution of growth function associated with solute concentration $c'_{a,m}$ in relation to the growth function of solvent concentration $c'$. We present a linear stability analysis, which carefully controls the unsteady base-state in such a way that the perturbations and the base-state vary with time simultaneously (see Appendix \ref{sec:Algo} for the algorithm used). For this purpose, Eqs. \eqref{eq:LP1}-\eqref{eq:LP3} are solved using a highly accurate pseudo-spectral method to convert the linearized perturbed equations into a system of ordinary differential equations with algebraic constraints.
 We apply discrete Fourier transform to all the unknown variables,
\begin{eqnarray}
\label{eq:DFT1}
& & c'(x,y,t) = \sum_{p,q}\hat{c}_{p,q}(t)e^{i(k_px + k_qy)}, \\
\label{eq:DFT2}
& & c'_{a,m}(x,y,t) = \sum_{p,q}\hat{c_1}_{p,q}(t)e^{i(k_px + k_qy)}, \\
\label{eq:DFT3}
& & \psi'(x,y,t) = \sum_{p,q}\hat{\psi}_{p,q}(t)e^{i(k_px + k_qy)}.
\end{eqnarray}
We also consider the discrete Fourier transform of the multiplicative terms,
\begin{eqnarray}
\label{eq:DFT4}
& & N(x,y,t) = \frac{\partial c^b}{\partial x} \frac{\partial \psi'}{\partial x} = \sum_{p,q}\hat{N}_{p,q}(t)e^{i(k_px + k_qy)}, \\
\label{eq:DFT5}
& & J(x,y,t) = \frac{\partial c^b}{\partial x} \frac{\partial \psi'}{\partial y} = \sum_{p,q}\hat{J}_{p,q}(t)e^{i(k_px + k_qy)}, \\
\label{eq:DFT6}
& & J_1(x,y,t) = \frac{\partial c_{a,m}^b}{\partial x} \frac{\partial \psi'}{\partial y} = \sum_{p,q}\hat{J_1}_{p,q}(t)e^{i(k_px + k_qy)},
\end{eqnarray}
where $k_p = 2\pi p/L_x, ~ k_q = 2\pi q/L_y, ~ p, q \in \mathbb{N}\cup\{0\}$. The coefficients of the Fourier transforms ($\hat{c}_{p,q},~ \hat{c_1}_{p,q}, ~ \hat{\psi}_{p,q}$, etc.) are computed using the fast Fourier transform (FFT) whenever $c'(x,y,t), ~ c'_{a,m}(x,y,t)$, $\psi'(x,y,t)$, etc. are known at the collocation points, $x_p = (p/N_x)L_x, ~ p = 0,1, \ldots, N_x-1$ and $y_q = (q/N_y)L_y, ~ q = 0,1, \ldots, N_y-1$. Here $N_x$ and $N_y$ correspond to the number of spectral points in the longitudinal and transverse directions, respectively. Substituting Eqs. \eqref{eq:DFT1}-\eqref{eq:DFT6} into Eqs. \eqref{eq:LP1}-\eqref{eq:LP3} following differential algebraic equations are obtained,
\begin{eqnarray}
\label{eq:DA1}
& & \frac{\mbox{d}\hat{c}_{p,q}}{\mbox{d}t} + \hat{J}_{p,q}= -\left(k_p^2 + k_q^2\right)\hat{c}_{p,q}, \\
\label{eq:DA2}
& & \frac{\mbox{d}\hat{c_1}_{p,q}}{\mbox{d}t} + \delta\hat{J_1}_{p,q} = -\left[i\lambda k_p + \delta\left(k_p^2 + k_q^2\right)\right]\hat{c_1}_{p,q}, \\
\label{eq:DA3}
& & \hat{\psi}_{p,q} = R\left(\hat{N}_{p,q} + ik_q\hat{c}_{p,q}\right)/\left(k_p^2 + k_q^2\right). 
\end{eqnarray}
Operator splitting method is employed to solve the differential equations (Eqs. \eqref{eq:DA1} and \eqref{eq:DA2}), subject to the algebraic constraints (Eq. \eqref{eq:DA3}). Using the values at the time level $t$, we predict $\hat{c}_{p,q}(t + \Delta t)$ and $\hat{c_1}_{p,q}(t + \Delta t)$ using the second order Adams-Bashforth method which are corrected using a trapezoidal rule. Inverse fast Fourier transform is used to obtain the corresponding values, $c'(x,y,t+\Delta t)$ and $c'_{a,m}(x,y,t+\Delta t)$, in the real space at the next time step $t + \Delta t$. Multiplicative terms are calculated in the real space at the new time level (the detailed algorithm of this numerical method can be found in Tan and Homsy \cite{Tan1988}). 

In order to use the benefit of the FFT we employ periodic boundary conditions on both the longitudinal and transverse boundaries, which are obtained straightforward from the physical boundary conditions, Eqs. \eqref{eq:perturb_BC1} and \eqref{eq:perturb_BC2}. Since, $x = 0, t = 0$ is a singular point for the error function base-state concentration profiles (Eqs. \eqref{concen_base_state} and \eqref{solute_base_state}), numerical simulations are performed by taking the initial time $t_0 = 10^{-3}$. Convergence study has been carried out by taking spatial discretization steps ($\triangle x$, $\triangle y$) = $(4, 8)$, and ($\triangle x$, $\triangle y$) = $(4, 4)$ in a computational domain $[0, 4096] \times [0, 512]$. Relative error has been computed using the standard Euclidean norm on $\mathbb{R}^2$ for amplification measure (see Eq. \eqref{eq:GR_LSA}) and it is found to be of $\mathcal{O}(10^{-2})$. To get optimal result, thus ($\triangle x$ , $\triangle y$) =(4, 4), with $\triangle t = 0.1 $ has been chosen. In the next section, the amplification measure and growth rate of the infinitesimal disturbances are presented.
\begin{figure}[!ht]
\centering
(a) \\
\includegraphics[width=3.5in, keepaspectratio=true, angle=0]{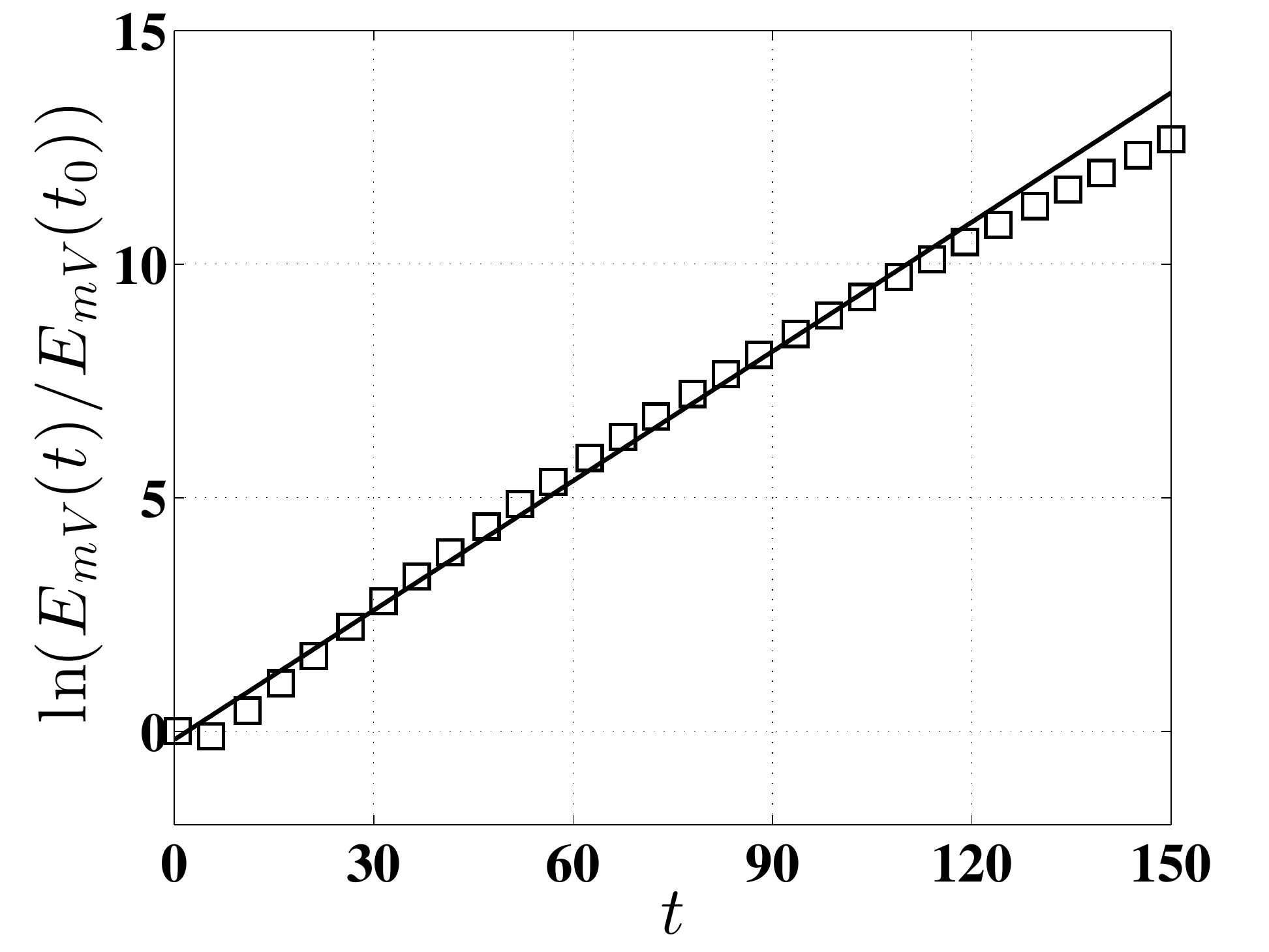} \\
(b) \\
\includegraphics[width=3.5in, keepaspectratio=true, angle=0]{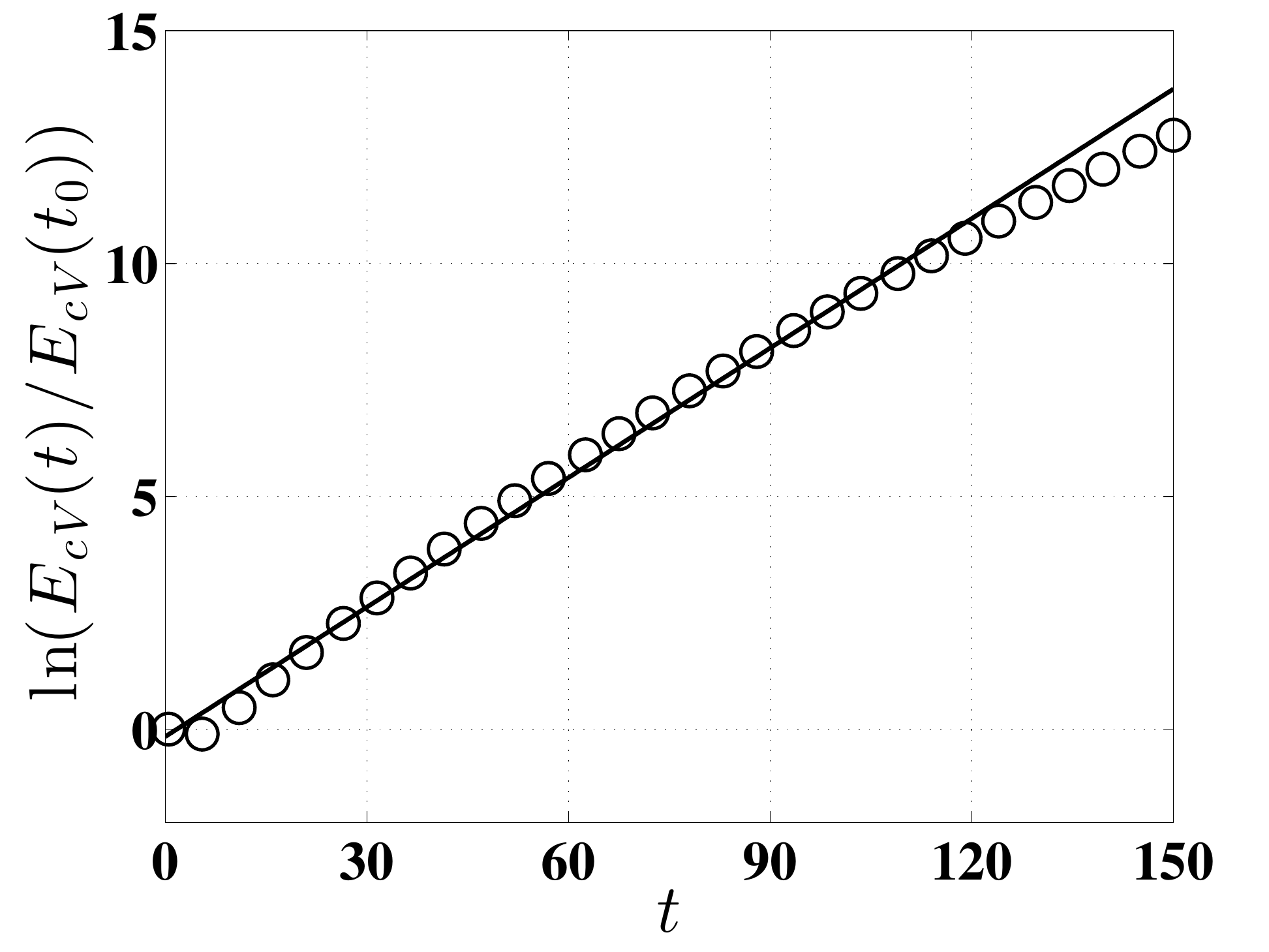}
\caption{A straight line fitting to the normalized amplification: (a) $\ln(E_{mV}(t)/E_{mV}(t_0))$ and (b) $\ln(E_{cV}(t)/E_{cV}(t_0))$ for $\kappa' = 0.2, R = 3, w = 512, k = 0.1$.  }\label{fig:growthrate_definition}
\end{figure} 

\subsection{Amplification and growth rate of the perturbations}\label{subsec:AGRP}
We use sinusoidal perturbations of the form, 
\begin{eqnarray}
\label{random_IC_Fourier}
g(x,y, t_0) = \begin{cases}
\epsilon*\cos(ky), & \text{at} ~ x = x_{r} ~ \text{and} ~ x_r + w\\
0, & \mbox{otherwise,}
\end{cases}
\end{eqnarray}
where $x_r$ is the position of the rear interface of the finite slice, $k$ is the non-dimensional wave number, $\epsilon$ is the amplitude of the perturbation, which is taken as $10^{-3}$ and $g$ corresponds to $c'$, $c'_{a,m}$, and $\psi'$. The present IVP approach measures the growth rate of each flow parameters individually which are obtained from amplification measures, without invoking QSSA. To quantify the amplification gain at time $t$, it is necessary to define a norm which is generally based on the kinetic energy of perturbations. We define 
\begin{eqnarray}
\label{eq:amp_c}
& & E_{cV}(t):= \int_{\Omega} (c'(x,y,t))^2  + (\vec{u'}(x,y,t))^2{\rm d}x{\rm d}y, \\
\label{eq:amp_cm}
& & E_{mV}(t):= \int_{\Omega} (c'_{a,m}(x,y,t))^2 +  (\vec{u'}(x,y,t))^2{\rm d}x{\rm d}y,
\end{eqnarray}
where $\Omega$ is the computational domain. $E_{cV}$ and $E_{mV}$ correspond to the amplification measures associated with the solvent and solute concentration perturbations coupled with the velocity perturbations, respectively. Similarly we can define, $E_c(t):= \int_{\Omega} (c'(x,y,t))^2 \text{\rm d}A$ and $E_m(t):= \int_{\Omega} (c'_{a,m}(x,y,t))^2 {\rm d}A$ to quantify the amplification measure of the solvent and solute concentration perturbations individually. 

The temporal evolution of the logarithm of the normalized amplification measures $E_{cV}(t)$ and $E_{mV}(t)$ are shown in Fig. \eqref{fig:growthrate_definition} for $R = 3, w = 512, k = 0.1$ and $\kappa' = 0.2$. It is observed that both $\ln(E_{cV}(t)/E_{cV}(t_0))$ (see Fig. \ref{fig:growthrate_definition}(a)) and $\ln(E_{mV}(t)/E_{mV}(t_0))$ (see Fig. \ref{fig:growthrate_definition}(b)) increase linearly at early times, determining an exponential growth of the perturbations. Thus, following Kumar and Homsy \cite{Kumar1999} the growth rate $\sigma_f(t)$ can be defined as follows. 

\begin{defn}[Growth rate and growth function]
The instantaneous growth rate of the perturbations is defined as,
\begin{equation}\label{eq:GR_LSA}
\sigma_f(t) \equiv \frac{1}{2}\frac{\mbox{d}[\ln(E_f(t))]}{\mbox{d}t},
\end{equation}
and the growth function is defined as the product of the instantaneous growth rate and the the corresponding time, i.e., $\sigma_f(t) t$, where $f \in \{ c, m, cV, mV\}$.
\end{defn}

\subsection{Direct numerical simulations}\label{subsec:DNS}
In this section we discuss direct numerical simulations of the perturbation equations using the pseudo-spectral method \cite{Tan1988}. Unlike LSA, in this case a finite amplitude perturbation is allowed by substituting $\theta = 1$ in Eqs. \eqref{eq:NLP1}-\eqref{eq:NLP3}. The resultant equations are solved following the same algorithm discussed in Sec. \ref{subsec:IVP_FSM}. The nonlinear (multiplicative) terms are defined as, 
\begin{eqnarray}
\label{eq:DFT7}
& & N(x,y,t) = \frac{\partial c^b}{\partial x}\frac{\partial \psi'}{\partial x} + \frac{\partial c'}{\partial x}\frac{\partial \psi'}{\partial x} + \frac{\partial c'}{\partial y}\frac{\partial \psi'}{\partial y} \nonumber \\
& & ~~~~~~~~~~~~ = \sum_{p,q}\hat{N}_{p,q}(t)e^{i(k_px + k_qy)}, \\
\label{eq:DFT8}
& & J(x,y,t) = \frac{\partial c^b}{\partial x}\frac{\partial \psi'}{\partial y} + \frac{\partial c'}{\partial x}\frac{\partial \psi'}{\partial y} - \frac{\partial c'}{\partial y}\frac{\partial \psi'}{\partial x} \nonumber \\
& & ~~~~~~~~~~~~ = \sum_{p,q}\hat{J}_{p,q}(t)e^{i(k_px + k_qy)}, \\
\label{eq:DFT9}
& & J_1(x,y,t) = \frac{\partial c_{a,m}^b}{\partial x}\frac{\partial \psi'}{\partial y} + \frac{\partial c'_{a,m}}{\partial x}\frac{\partial \psi'}{\partial y} - \frac{\partial c'_{a,m}}{\partial y}\frac{\partial \psi'}{\partial x} \nonumber \\
& & ~~~~~~~~~~~~ = \sum_{p,q}\hat{J_1}_{p,q}(t)e^{i(k_px + k_qy)}. 
\end{eqnarray}
For the validation of the numerical code the results of Mishra \textit{et al.} \cite{Mishra2009} are reproduced. Further, the growth rate obtained from DNS is calculated from Eq. \eqref{eq:GR_LSA}. In the absence of adsorption, i.e., $\kappa ' =0$, the solute remains in the mobile phase and follows the dynamics of the sample solvent and acts like a passive scalar. Thus, the growth function $\sigma_{mV} t$ associated with the solute concentration must be the same as that of the solvent concentration $\sigma_{cV} t$ (see Eqs. \eqref{eq:NLP1} and \eqref{eq:NLP3}). 


\section{Stability analysis without adsorption}\label{sec:NoAdsorp}

\subsection{Displacement of two semi-infinite fluids}\label{subsec:single}
In this section, the linear stability analysis of classical VF instability between two miscible fluids in a porous medium is revisited in the absence of adsorption, i.e., when $\kappa'=0$. It is observed that  letting $w \to \infty$ in Eqs. \eqref{concen_base_state} we obtain $c^b(x,t) = 1/2[1 + \mbox{erf}(x/2\sqrt{t})]$, which corresponds to the stability problem associated with the displacement of two semi-infinite fluids \cite{Tan1986}. In the framework of the present formulation this can be obtained by considering the sample width $w$  very large in Eqs. \eqref{concen_base_state} and \eqref{random_IC_Fourier}, so that the right interface remains unaffected from the dynamics at the left interface. Corresponding stability equations are solved using the IVP method described in Sec. \ref{subsec:IVP_FSM}. 

\begin{figure}[!ht]
\centering
\includegraphics[width=3.5in, keepaspectratio=true, angle=0]{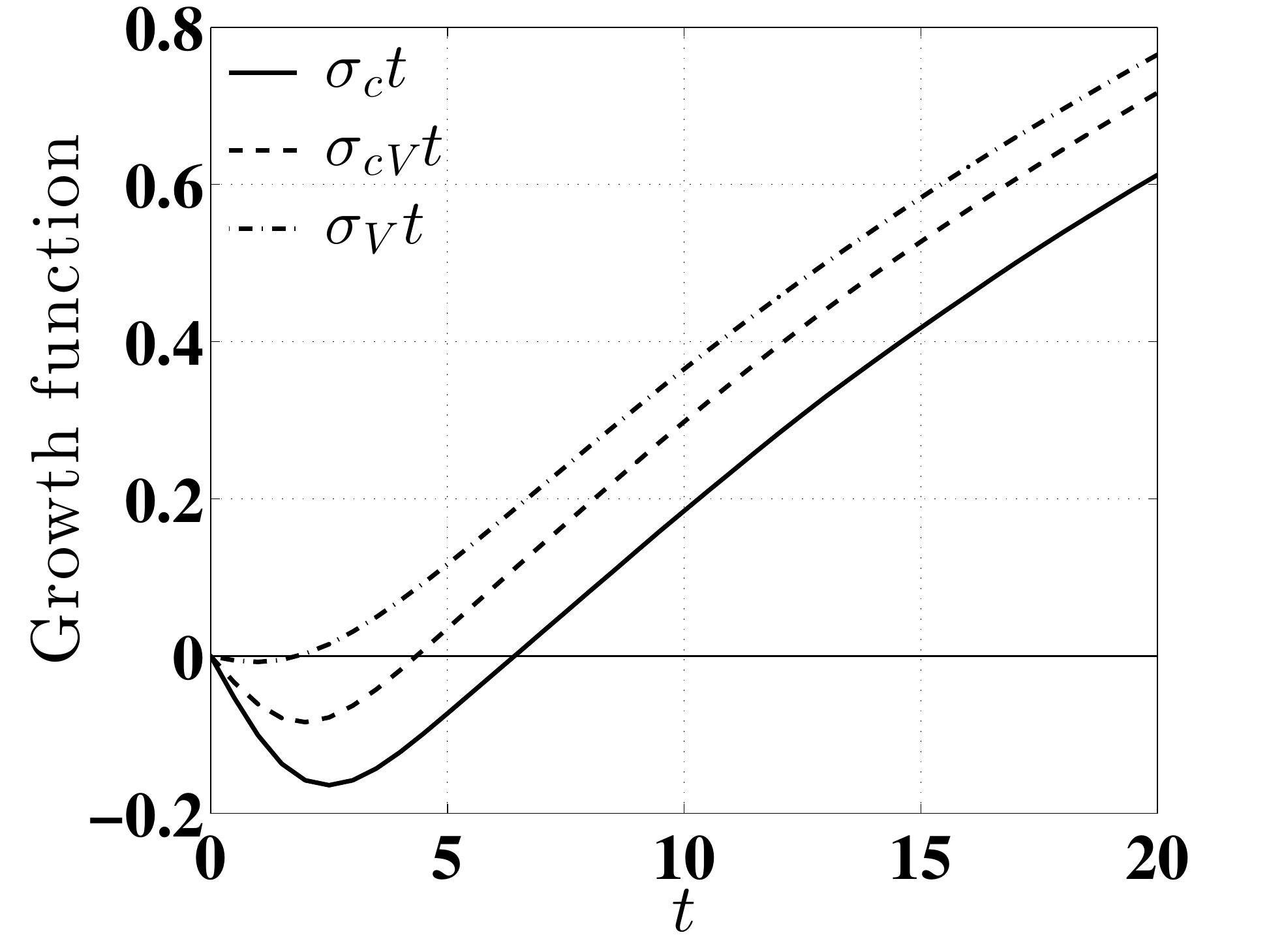}
\caption{Temporal evolution of the growth functions for two semi-infinite fluids with $R = 2, k = 0.1$. }\label{fig:growth_single} 
\end{figure} 

Following the definition in Eq. \eqref{eq:GR_LSA}, growth functions, $\sigma_{c} t, \sigma_{V} t$ and $\sigma_{cV} t$ are calculated, where $\sigma_{V}$ is measured from $E_V = E_{cV} - E_c$. The temporal evolution of these growth functions are shown in Fig. \ref{fig:growth_single} for $R = 2, k = 0.1$. This figure depicts that all the disturbances initially decay in diffusion dominated regime and grow when convection dominates, and eventually become unstable at a later time.  It is observed that for all $t > t_0$, $\sigma_ct < \sigma_{cV}t < \sigma_Vt$,  which is consistent with the results of Tan and Homsy \cite{Tan1986}, who discussed only $\sigma_c$ and $\sigma_{V}$ in their study. Further, it is verified that perturbations to each flow variables are essential in hydrodynamic stability analysis. We observe that if perturbation is not imposed to $\psi'$, $\sigma_{V}$ becomes unconditionally unstable for all $t > t_0$, which violates the physics of early time diffusion dominated regime. As the velocity plays an important role in VF instability, the growth function $\sigma_{cV}t$ is used to analyze the onset of instability for the classical VF problem and also for finite slice displacement (see Sec. \ref{subsec:finite}). The onset of instability is defined as,
\begin{equation}
t^c = \min_{\tau}\left \{ \tau > t_0 : \sigma_{cV}(t \leq \tau) \leq 0, ~ \sigma_{cV}(t > \tau) > 0 \right\}. 
\end{equation} 

In summary, the present LSA method differs from the other existing linear stability analyses by two conceptual distinctions. It successfully captures both the onset of instability and the early time diffusion dominated regime, which was never achieved with the well known QSSA methods \cite{Tan1986, Kim2012, Pramanik2013}. Also, the present LSA measures the magnitude of the growth function is close to zero at the early time, which represents that at such initial period the growth function is of order $\mathcal{O}(1)$ \cite{Tan1986}. 

\begin{figure}[!ht]
\centering
(a) \\
\includegraphics[width=3.5in, keepaspectratio=true, angle=0]{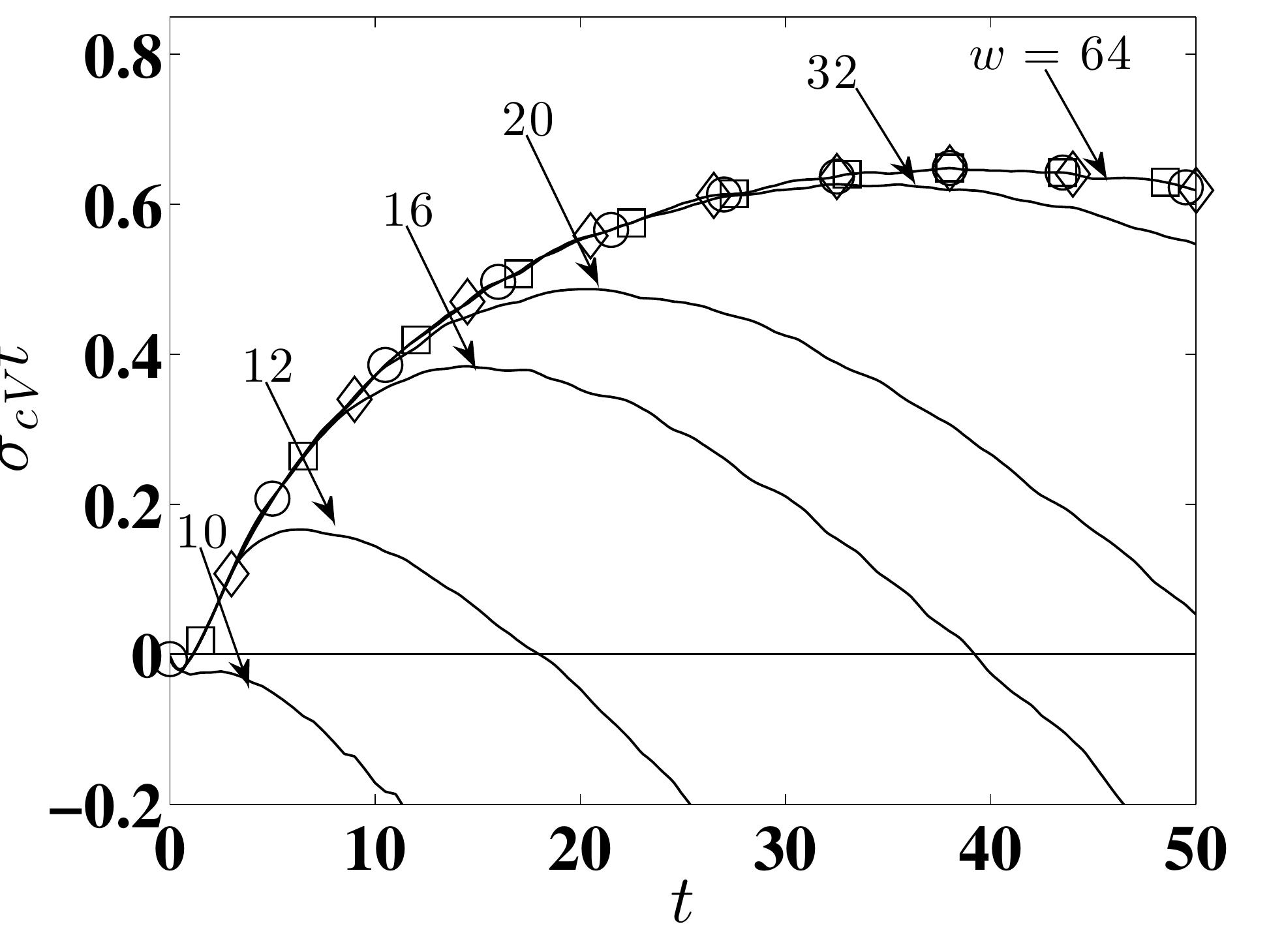} \\
(b) \\
\includegraphics[width=3.5in, keepaspectratio=true, angle=0]{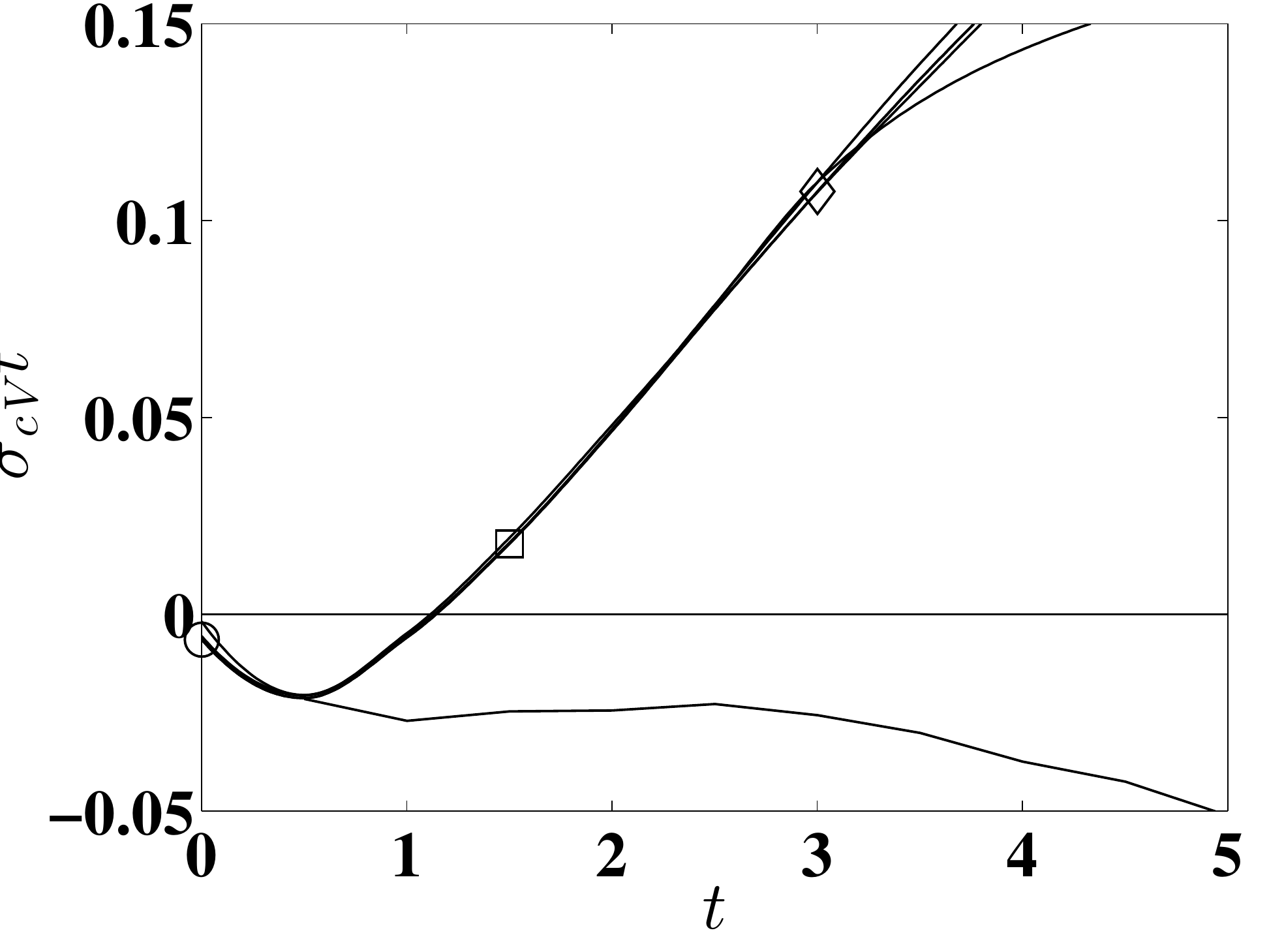}
\caption{(a) Temporal evolution of the growth function, $\sigma_{cV}t$, for $R = 2, k = 0.2$ and different width of the finite slice, $w$. Symbols correspond to growth function associated with $ w = 128 (\bigcirc), ~ 256 (\square)$ and $\infty (\lozenge)$. (b) Magnified of (a) near the onset of instability. }\label{fig:L_effect}
\end{figure}

\subsection{Finite slice displacement}\label{subsec:finite}
Here we are interested to investigate the influence of the finite slice width $w$ on the onset of VF dynamics between the displacing fluid and the sample solvent. Fig. \ref{fig:L_effect} depicts the temporal evolution of $\sigma_{cV}t$ for $R = 2, k = 0.2$ and different sample width $w$. For these parameters the critical sample width for the onset of instability has been found to be $w_c \approx 10$ (see Fig. \ref{fig:L_effect}). It is shown that the onset of instability are indistinguishable for all $w \geq w_c$ (see Fig. \ref{fig:L_effect}(b)). Further, we observe that the growth functions corresponding to a finite slice of width $w \gtrsim 64$ (see Fig. \ref{fig:L_effect}(a)) and the displacement of two semi-infinite fluids are indistinct. This result is consistent with the linear stability results of Pramanik and Mishra \cite{Pramanik2013} and numerical simulations of De Wit {\it et al.} \cite{Ann2005}. For the increasing values of $R$ the viscosity contrast increases which influences the disturbances to grow faster. Therefore, it can be shown that the critical sample width $w_c$ decreases as $R$ increases \cite{Pramanik2013}.

\begin{figure}[!ht]
\centering
\includegraphics[width=3.5in, keepaspectratio=true, angle=0]{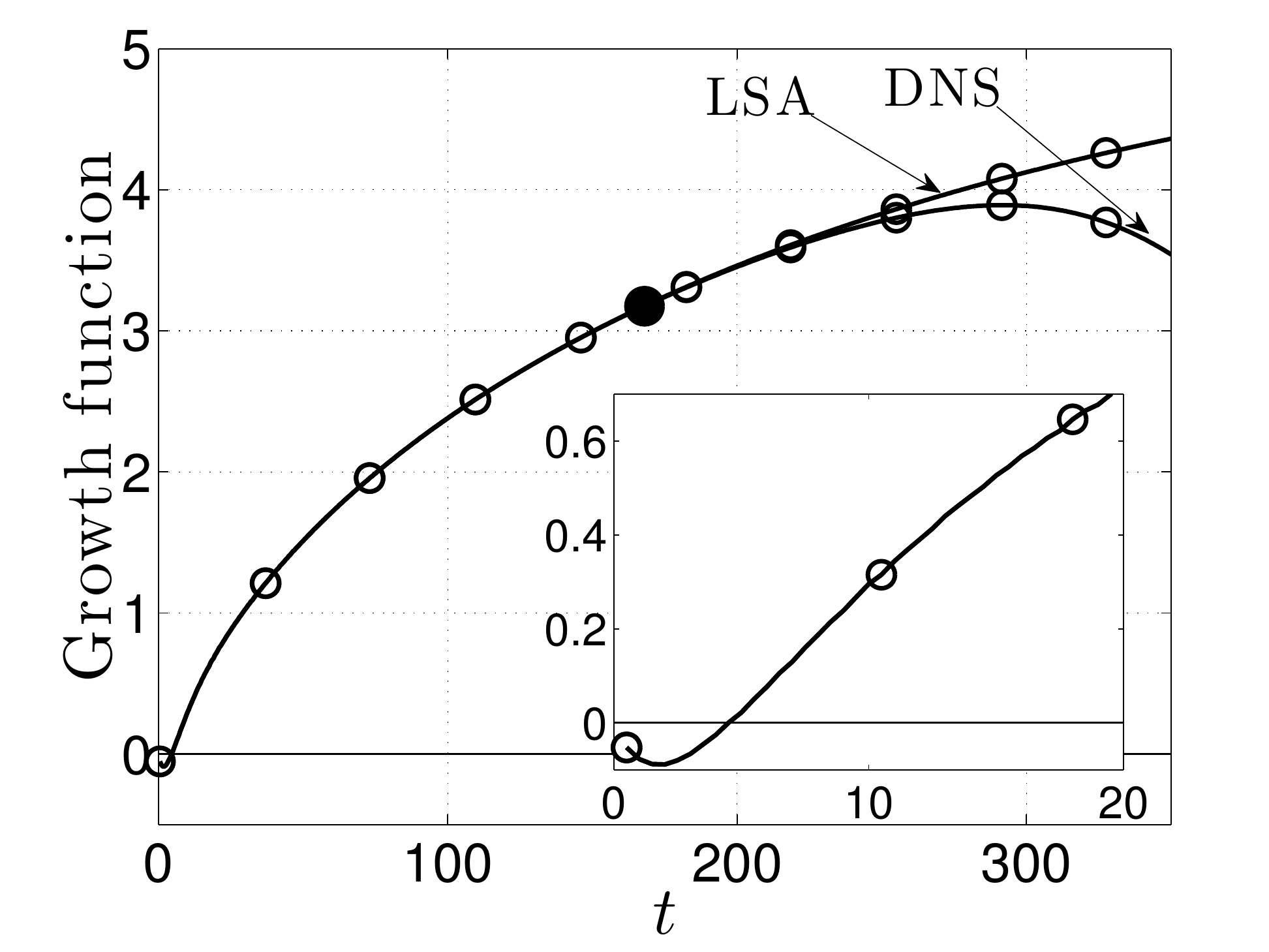}
\caption{Growth functions, $\sigma_{cV} t$ (\solidrule) and $\sigma_{mV}t$ ($\bigcirc$), for $\kappa' = 0, R = 2, w = 512, k = 0.15$. The solid dot ($\CIRCLE$) represents the onset of the nonlinearity. Inset: Magnified near the onset of instability. }\label{fig:growth_kappa_0} 
\end{figure} 

In the absence of adsorption of the solute on the porous matrix, the present model reduces to the model of De Wit {\it et al.} \cite{Ann2005}, and this was confirmed by Mishra {\it et al.} \cite{Mishra2009} through DNS. Before we analyse the influence of retention parameter, $\kappa'$, on the stability of the adsorbed solute, we confirm the same through the present LSA. In this context we choose $R = 2, k = 0.15, \kappa' = 0$, and compare the temporal evolution of the growth functions, $\sigma_{cV}t$ and $\sigma_{mV}t$. These growth functions are shown in Fig. \ref{fig:growth_kappa_0}, which depicts that $\sigma_{cV} t$ and $\sigma_{mV} t$ are indistinguishable, thus confirming the validity of the present IVP based LSA. Further, the growth rate obtained from DNS is compared with those obtained from LSA in Fig. \ref{fig:growth_kappa_0}. It shows that the DNS results coincide with the LSA results before non-linearity becomes dominant at $t \approx 167.5$. Therefore, linear stability theory is not valid beyond this time and one needs to solve the complete nonlinear problem to capture the dynamics of the instability pattern. 

\section{Stability analysis of an adsorbed solute}\label{sec:Adsorption}
In this section the numerical results obtained from both LSA and DNS are discussed to analyze the effect of the retention parameter, $\kappa'$, on the growth of the perturbations associated with the  solute dynamics. As $\kappa'$ has no effect on the temporal evolution of the sample solvent $c$, we choose the growth function $\sigma_{mV}t$ associated with the solute concentration $c_{a,m}$ to quantify the effect of linear adsorption isotherm on solute dynamics.

\begin{figure}[!ht]
\centering
\includegraphics[width=3.5in, keepaspectratio=true, angle=0]{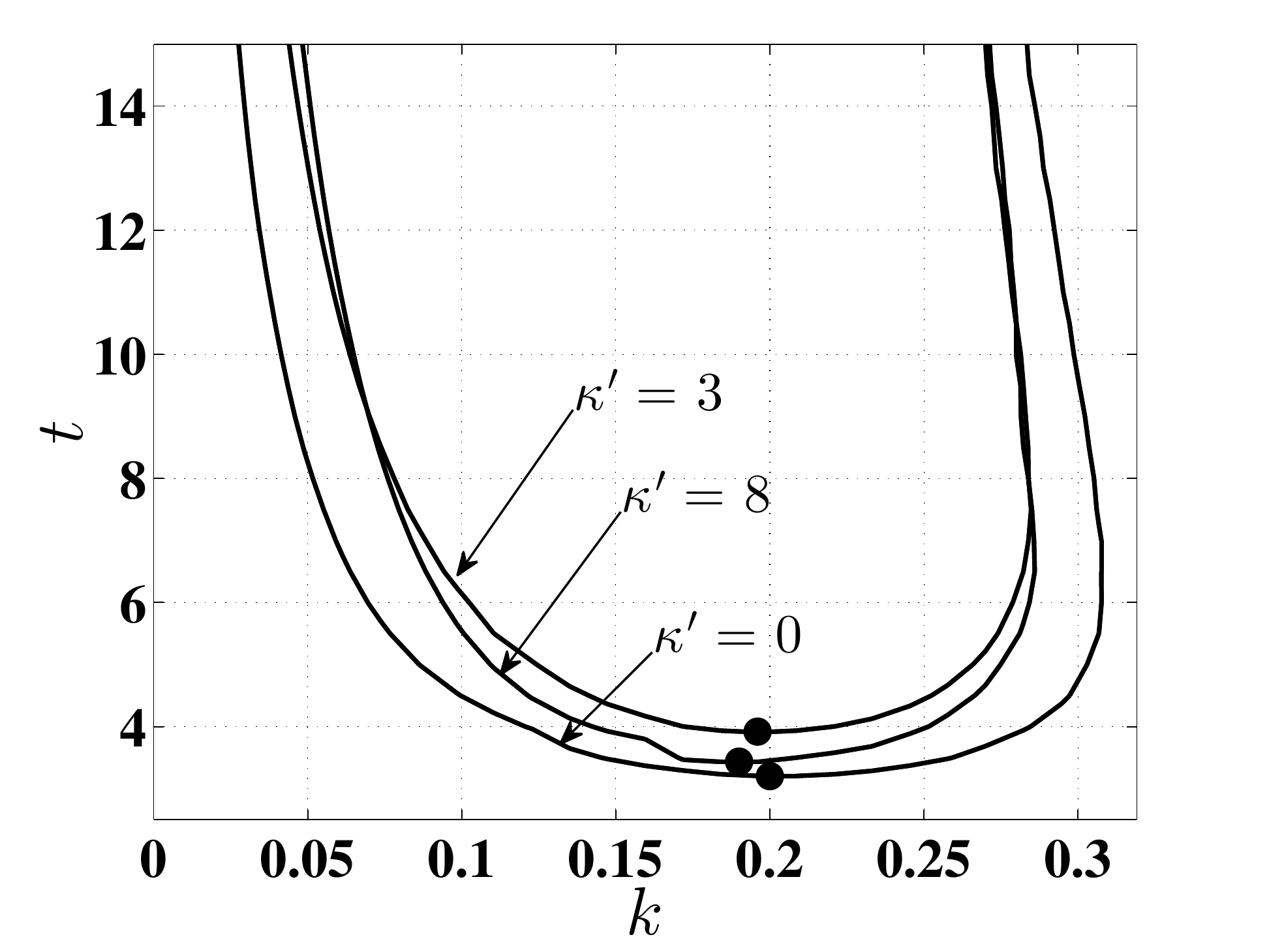} 
\caption{Neutral curves for $R = 2, w = 512$ and different values of the retention parameter $\kappa' = 0, 3$ and $8$. The critical time $t^c$ and corresponding wave number $k^c$ on the parameter space is marked by solid dot ($\CIRCLE$). }\label{fig:neutral}
\end{figure}

\subsection{Neutral curves and dispersion curves}\label{subsec:NCDC}
The overall characteristic of the temporal instability driven by viscous forces can be compactly presented in a phase space spanned by the perturbation wave number $k$ and time $t$. Fig. \ref{fig:neutral} represents the isocontours of the growth rate, $\sigma_{mV}$ = 0, in the ($k,t$) parameter plane for $R = 2, w = 512$ and the retention parameter $\kappa' = 0, 3$ and $8$. The area above each contour line corresponding to different $\kappa'$ represent the unstable region for the respective values of the retention parameter. For small times all perturbation wave numbers are stable, later a band of wave numbers become unstable. The critical point $(k^c, t^c)$ is marked with a solid dot at the lowest point on the isocontours $\sigma_{mV} = 0$. We show that the region of instability is  larger for non-adsorbed case ($\kappa' = 0$) in comparison to the case when $\kappa' \neq 0$. This signifies that adsorption of the solute concentration on the porous matrix reduces the instability. Further, it is observed that the onset time $t^c$ has a non-monotonic dependence on $\kappa'$. The detail analysis of this non-monotonicity and the underlying physics are discussed in Sec. \ref{subsec:retention_onset}.

\begin{figure}[h]
\centering
\includegraphics[width=5in, keepaspectratio=true, angle=0]
{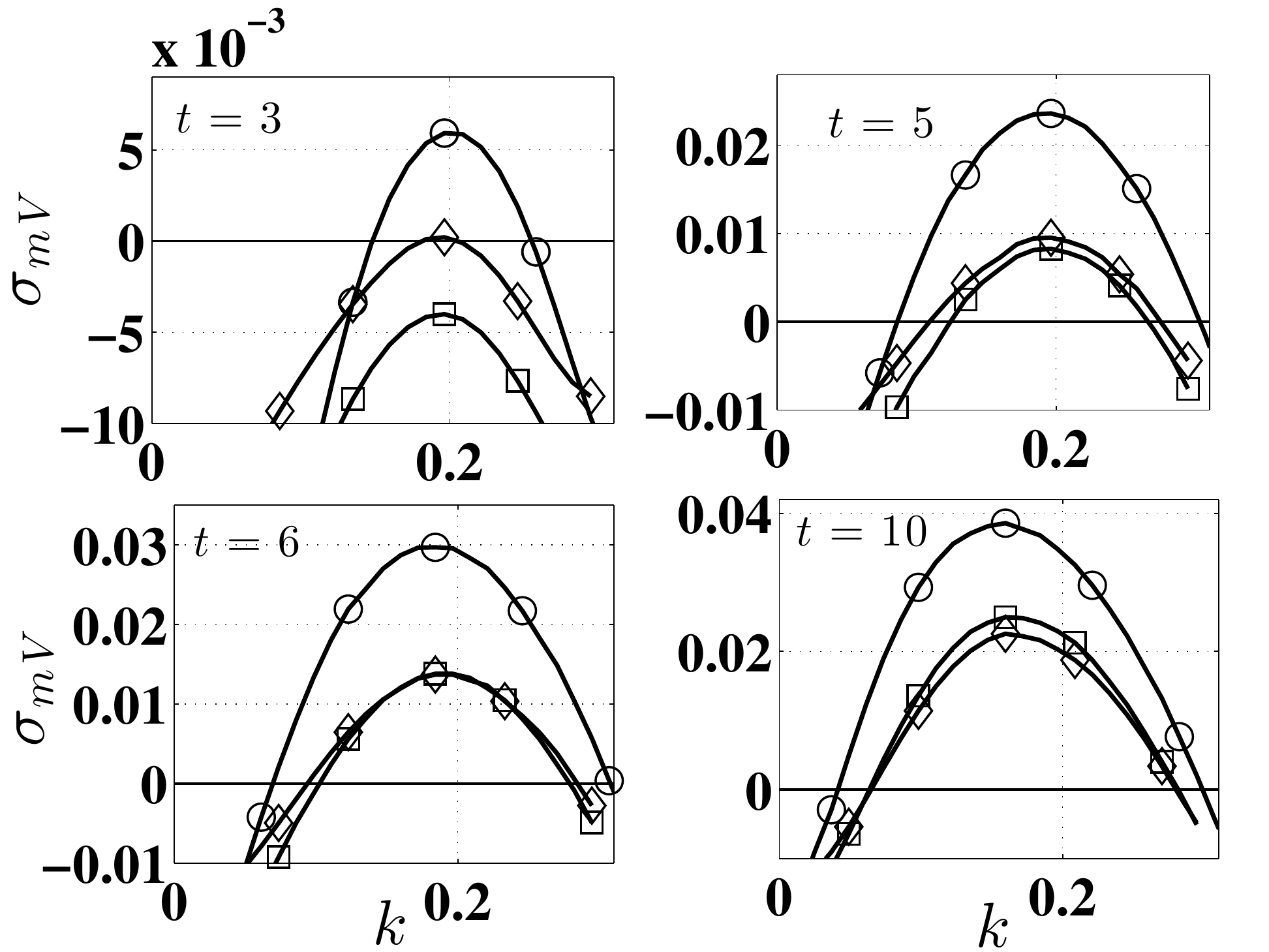}
\caption{Dispersion curves at different times for $w = 512, R = 2$: $\kappa' = 0 (\bigcirc), 3 (\square), 8 (\lozenge)$. }\label{fig:dispersion_curves_R_2_L_512_new}
\end{figure} 

Dispersion relation is a convenient measure of the growth rate corresponding to the different wave numbers of the disturbances. Fig. \ref{fig:dispersion_curves_R_2_L_512_new} depicts the dispersion curves corresponding to the same parameters as in Fig. \ref{fig:neutral} at different times. This figure shows that adsorption of the solute on the porous matrix has an overall stabilizing influence. Physically, this illustrates the fact that fingers which are too narrow are immediately smoothed out by the increase in transverse dispersion. Not only the growth rate of the perturbation associated with certain wave number is reduced with increasing $\kappa'$, the spectrum of the unstable wave numbers is also reduced as the retention parameter is changed from $\kappa' = 0 $ to $\kappa' \neq 0$. Besides showing the spectrum of unstable wave numbers, dispersion curves also represents the magnitude of the growth rate of the unstable modes. At a given time the maximum possible instantaneous growth rate that can be achieved by any initial condition is  denoted by $\sigma_{mV}^{\rm max}(t)$ and is defined as,
\begin{equation}\label{eq:sigma_max}
\sigma_{mV}^{\rm max}(t) = \displaystyle\sup_{0 \leq k < \infty} \sigma_{mV}(k,t),
\end{equation}
and the corresponding wave number is the most unstable wave number, which is denoted by $k_{\rm max}(t)$, at time $t$. Fig. \ref{fig:dispersion_curves_R_2_L_512_new} shows that for $R = 2, w = 512$ and $t \lesssim 10$ the most unstable wave number, $k_{\rm max}$, which decreases with $t$, is almost independent of the retention parameter, $\kappa'$ (see the corresponding curves for $\kappa' =0, 3, 8$). Corresponding $\sigma_{mV}^{\rm max}$ increases with $t$ having a non-monotonic dependence on the retention parameter $\kappa'$, for $t \lesssim 6$. 

\begin{figure}[h]
\centering
\includegraphics[width=3.5in, keepaspectratio=true, angle=0]
{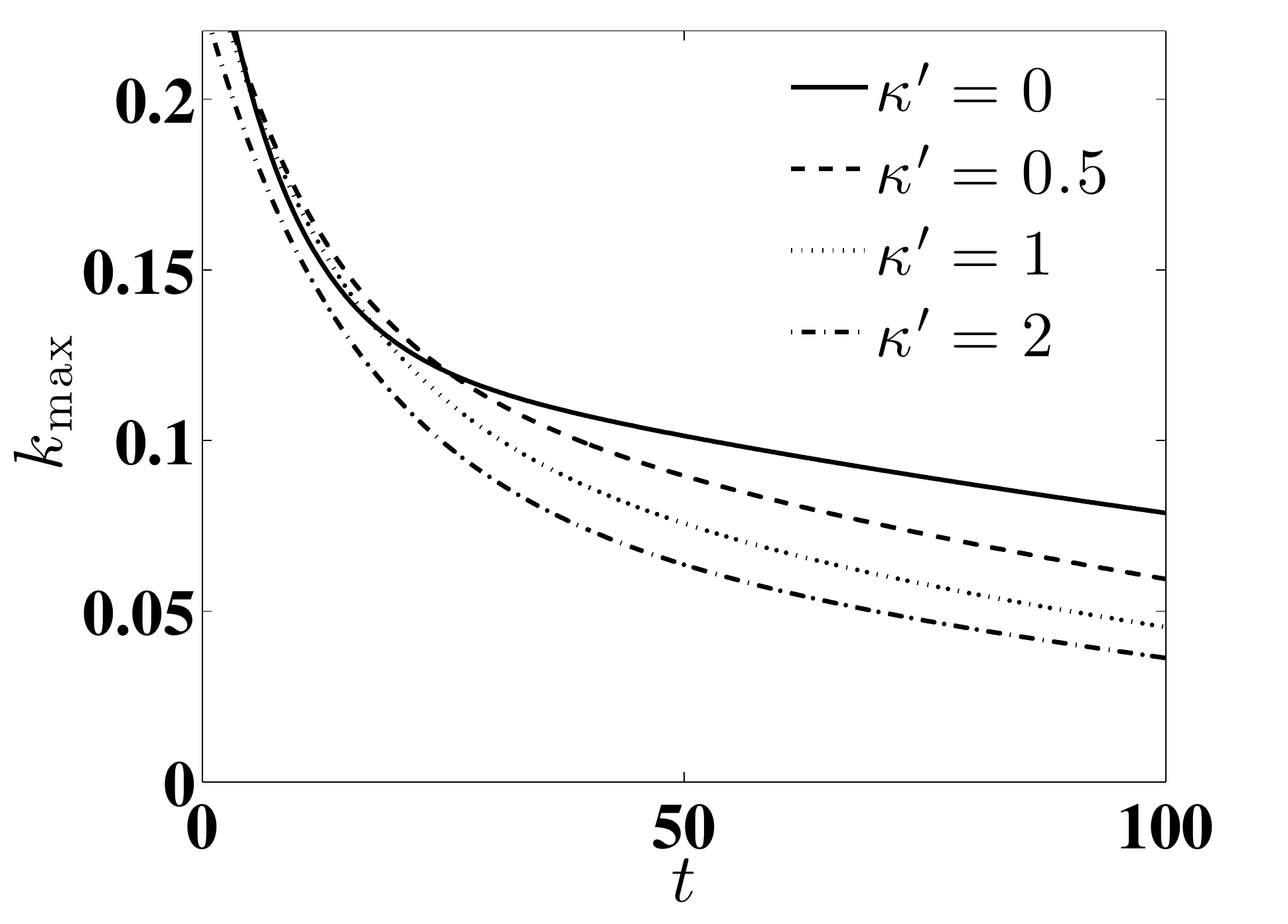}
\caption{Dominant wave number, $k_{\rm max}$, of the solute concentration for $w = 512, R=2$ and different values of $\kappa'$. }\label{fig:dominant_waveno}
\end{figure}

In Fig. \ref{fig:dominant_waveno} the temporal evolution of the most unstable wave numbers are presented for $R = 2, w = 512$ and the retention parameter, $\kappa' = 0, 0.5, 1$ and $2$. This figure depicts that for all $\kappa'$, $k_{\rm max}$ decreases monotonically with $t$, indicating stabilization of short wave perturbations at later times. We show that, $k_{\rm max}$ is almost independent of $\kappa'$ at early times, while at later times $k_{\rm max}$ decreases monotonically when $\kappa'$ increases, which signifies that the long wave perturbations are smoothed out more rapidly with increasing $\kappa'$.

\begin{figure}[h]
\centering
\includegraphics[width=3.5in, keepaspectratio=true, angle=0]{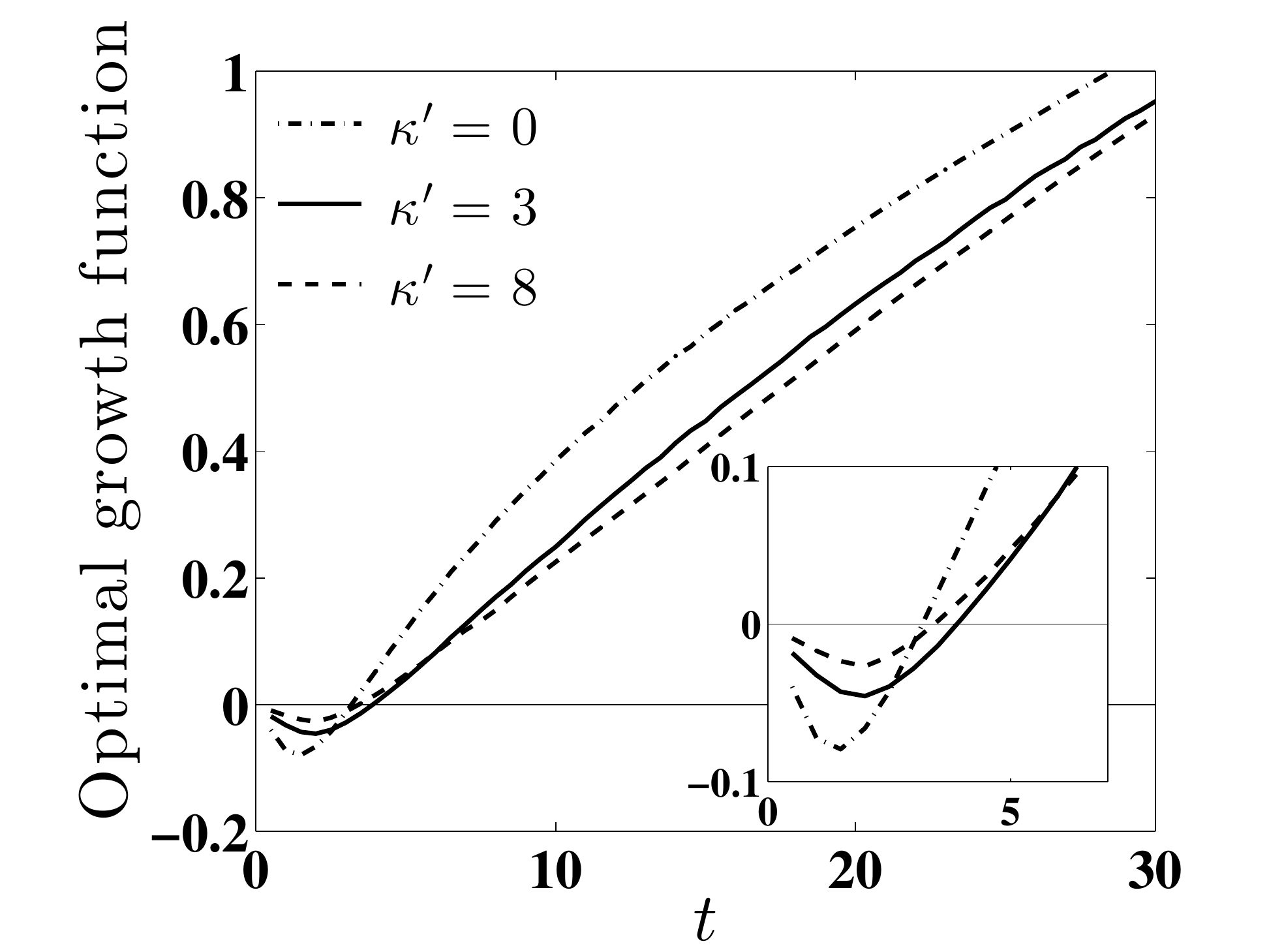}
\caption{Optimal growth function, $\sigma_{mV}^{\rm max}t$, of the solute concentration for $w = 512, R=2$ and different values of $\kappa' = 0, 3, 8$. Inset shows the non-monotonicity of onset time and dominance of diffusion at early times.}\label{fig:optimal_growth_1}
\end{figure} 

\subsection{Optimal growth}\label{subset:OG}
In an experiment or a real physical system a perturbation consists of combination of different wave numbers, so it is important to  calculate the onset of instability considering all possible wave numbers. Fig. \ref{fig:optimal_growth_1} represents $\sigma_{mV}^{\rm max} t$ for $R = 2, w = 512$ and $\kappa' = 0, 3, 8$. It confirms that adsorption of the solute on the porous matrix results into an overall delay of the onset of instability. The qualitative as well as quantitative effect of $\kappa'$ on the onset of instability are discussed below. It must be noted that the retention parameter has no effect on the dynamics of the sample solvent. From the mass-balance equation for solute (see Eq.\eqref{solute}) it is seen that the dispersion coefficient is $1/(1+\kappa')$, while the travelling wave velocity of the mobile phase solute concentration is $-\kappa'/(1 + \kappa')$ in the longitudinal direction. Therefore, as $\kappa'$ increases the solute moves in the upstream direction away from the sample solvent more rapidly and eventually disengages from the solvent zone. In this process the rear or the frontal interface of the solute zone features VF instability depending on the value of the retention parameter, $\kappa'$ \cite{Mishra2009}. As $\kappa'$ increases, from Eq.\eqref{solute} it can be observed that the dispersion coefficient decreases and hence the optimal growth of disturbances for $\kappa' = 0$ damped most in comparison to $\kappa' \neq 0$. This is depicted in the inset of Fig. \ref{fig:optimal_growth_1}. Further, for $\kappa' >3$, the onset time is early in comparison to when $\kappa' \in (0,3]$. Thus the onset time varies non-monotonically with respect to $\kappa'$. 

\begin{figure}[!ht]
\centering
\includegraphics[width=3.5in, keepaspectratio=true, angle=0]{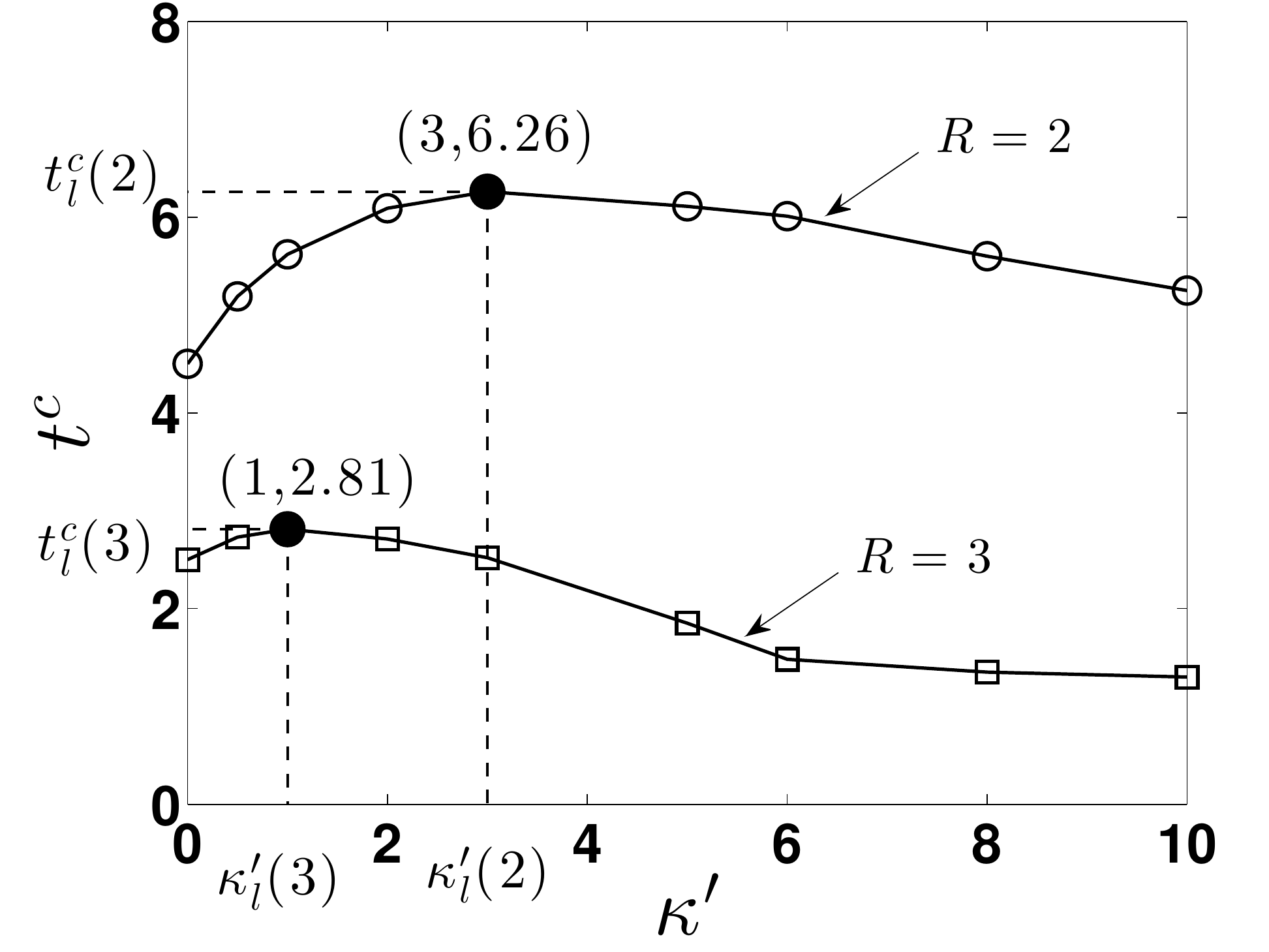}
\caption{Onset time, $t^c$ calculated from $\sigma_{mV}$, as a function of the retention parameter, $\kappa'$, for $R = 2, 3, w = 512, k = 0.1$. }\label{fig:onsetLSA}
\end{figure}

\subsection{Effect of retention parameter and log-mobility ratio on the onset of instability}\label{subsec:retention_onset}
In order to quantify the influence of the retention parameter $\kappa'$ on the onset of instability, we solve the IVP, Eqs. \eqref{eq:LP1}-\eqref{eq:LP3}, for two different log-mobility ratio $R = 2$ and $3$, sample width $w = 512$ and $\kappa' \in (0,\infty)$. Thus calculated onset time $t^c$ is shown as a function of $\kappa'$ in Fig. \ref{fig:onsetLSA}. This figure represents the onset time calculated from $\sigma_{mV}$ and depicts that $t^c$ depends non-monotonically on $\kappa'$. In order to analyze the non-monotonic behavior of the onset of instability more precisely, for a given wave number $k$ and log-mobility ratio $R$ we define the largest onset time $t_{l}^c(\kappa'_{l}(R),k, R)$ over all retention parameter $\kappa'$ as,  
\begin{equation}\label{eq:optimal_retention}
t_{l}^c(\kappa'_{l},k, R) \equiv \max_{\kappa'} ~ t^c(k,R),
\end{equation}
and the corresponding retention parameter is denoted by $\kappa'_{l}(R)$. In particular, we choose $k = 0.1$ to compute $t^c_{l}$ and the corresponding $\kappa'_l(R)$ for two values of log-mobility ratio $R = 2$ and $3$. It is determined that $t^c_l(R = 2) = 6.26, \kappa'_l(R = 2) = 3$ and $t^c_l(R = 3) = 2.81, \kappa'_l(R = 3) = 1$ and they are represented by solid dots in Fig. \ref{fig:onsetLSA}. More generally, it can be shown that $R_1 < R_2$ implies $\kappa'_{l}(R_1) > \kappa'_{l}(R_2)$ and $t_{l}^c(\kappa'_{l}(R_1), R_1) - t^c(0, R_1) > t_{l}^c(\kappa'_{l}(R_2), R_2) - t^c(0, R_2)$. 

We further show that, for sufficiently large values of $\kappa'$, the instability in the adsorbed solute sets in earlier than the solvent. The corresponding value of $\kappa'$ depends on the log-mobility ratio, and it decreases with decreasing $R$. For large values of $\kappa'$ the axial dispersion of the solute concentration is smaller than that of the solvent concentration. Thus the stabilization effect of the dispersion is more on the perturbations of the solvent concentration in comparison to the solute concentration. Hence, in the early time diffusion dominated regime $\sigma_m$ decays slower compared to $\sigma_c$. As a consequence, $\sigma_{mV}$ becomes larger than $\sigma_{cV}$ and hence shows an early onset for the solute concentration than the solvent concentration.

\begin{figure}[!ht]
\centering
(a)\\
\includegraphics[width=3.5in, keepaspectratio=true, angle=0]{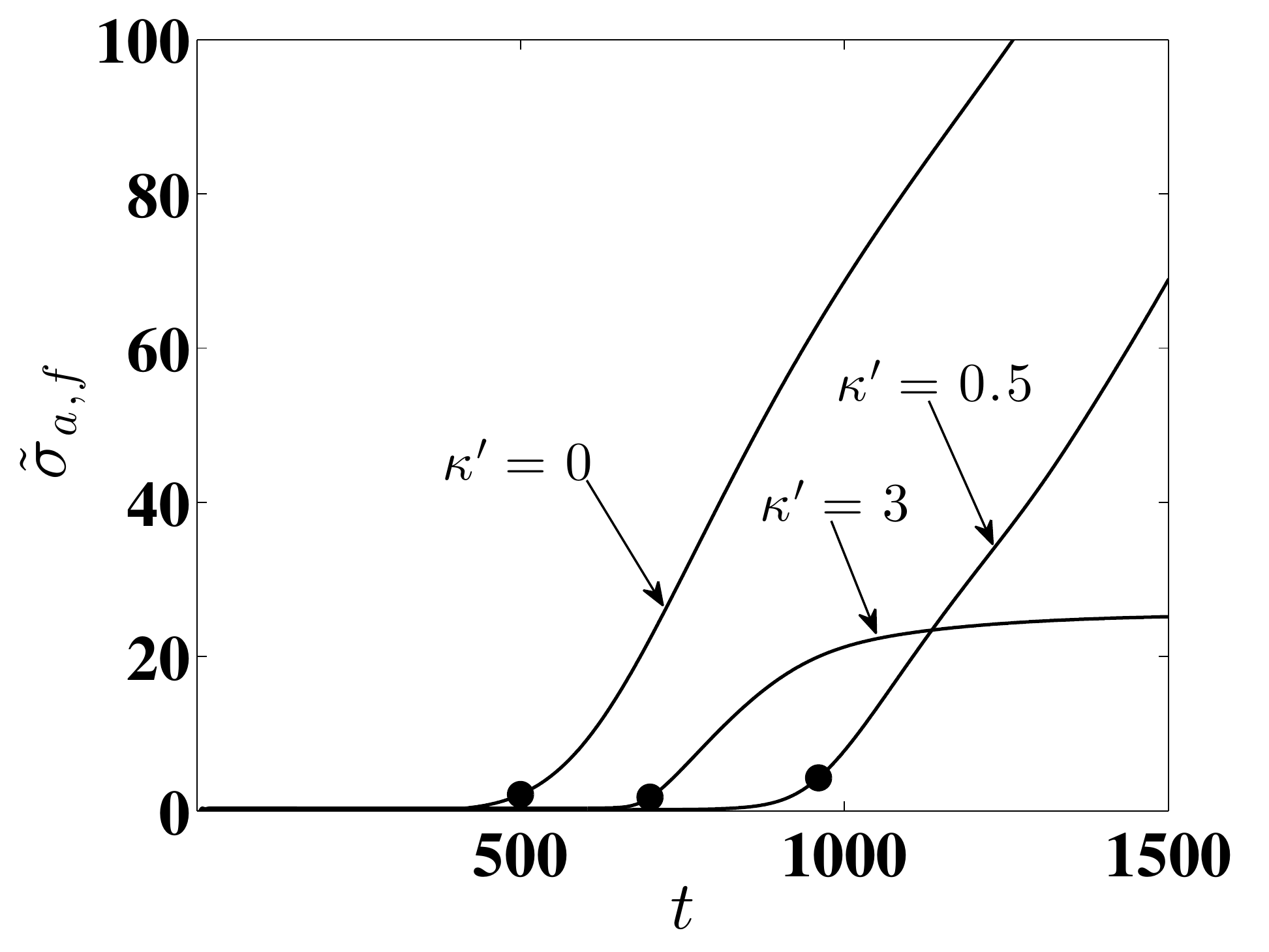}\\
(b)\\
\includegraphics[width=3.5in, keepaspectratio=true, angle=0]{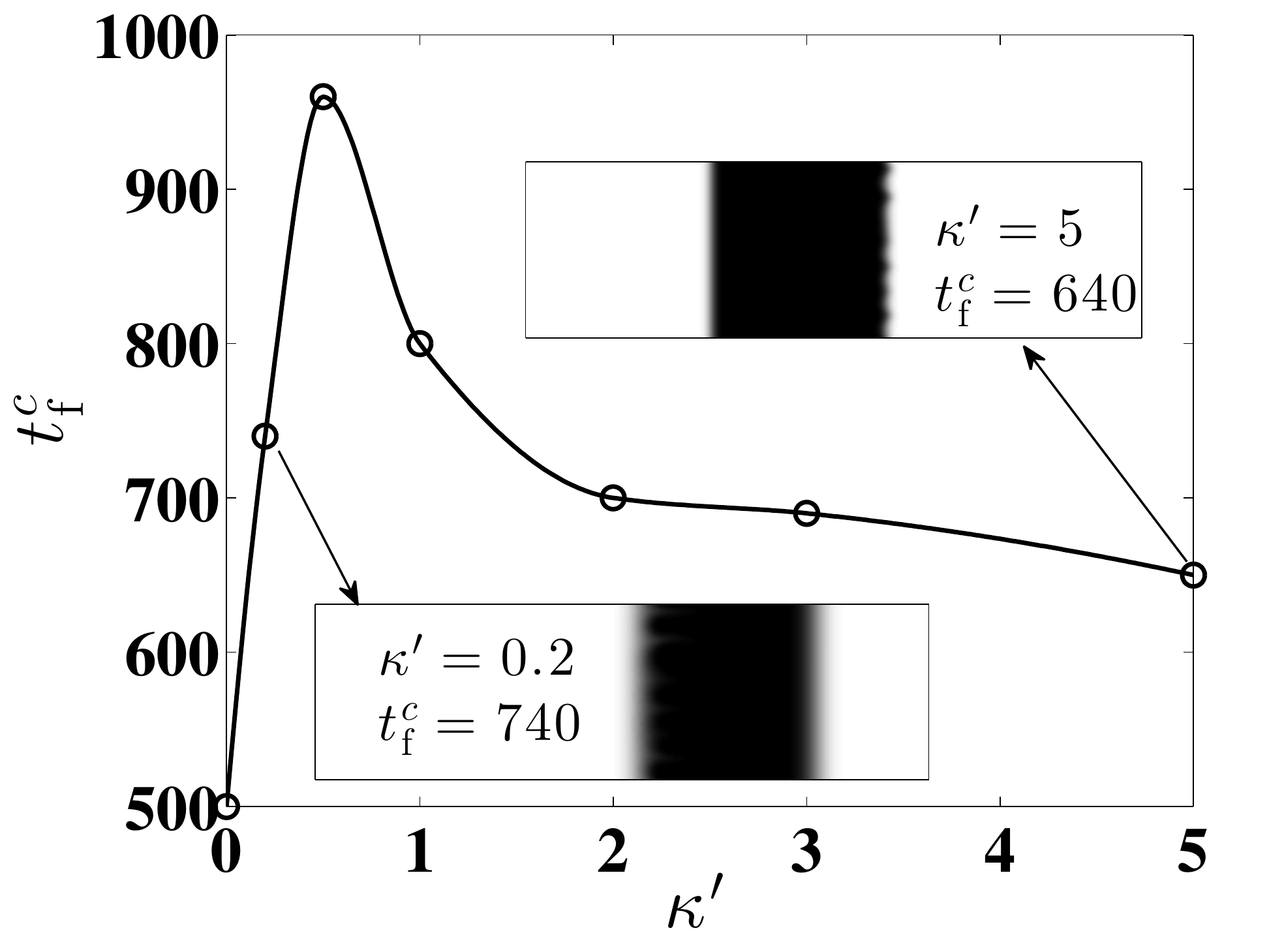}
\caption{For $R = 2, w = 512$: (a) Contribution of fingering to the standard deviation $\tilde{\sigma}_{a,f}$ as a function of time, (b) Onset time, $t^c_{\rm f}$, obtained from DNS as a function of the retention parameter, $\kappa'$. Inset: Density plot of the solute  concentration field for $\kappa' = 0.2$ and $\kappa' = 5$ at the corresponding onset time, $t^c_{\rm f} = 740$ and $t^c_{\rm f} = 640$, respectively. }\label{fig:onsetDNS}
\end{figure} 

Next we analyze the dependence of the onset of fingers in the nonlinear regime on $\kappa'$ by solving the fully nonlinear equations \cite{Ann2005, Mishra2009}. De Wit {\it et al.} \cite{Ann2005} showed that the standard deviation of the solvent concentration zone due to viscous fingering $\tilde{\sigma}_f$ starts growing from zero at the onset of VF. Further, following their analysis, Mishra {\it et al.} \cite{Mishra2009} measured 
\begin{equation}
\tilde{\sigma}_{a,f}^2(t) = \frac{\int_0^{L_x}x^2\bar{c}_{a,m}\;\mbox{d}x }{\int_0^{L_x}\bar{c}_{a,m}\;\mbox{d}x} -\left(\frac{\int_0^{L_x}x\bar{c}_{a,m}\;\mbox{d}x}{\int_0^{L_x}\bar{c}_{a,m}\;\mbox{d}x}\right)^2 - \left( \frac{w^2}{12} +\frac{2 t}{1+ \kappa'}\right), 
\end{equation}
for the solute concentration zone, where $\bar{c}_{a,m}(x,t)= \int_{0}^{L_y} c_{a,m}(x,y,t) \mbox{d}y$ is the transversely averaged solute concentration profile. For  $R=2, w=512$, Fig. \ref{fig:onsetDNS}(a) depicts the contribution to the standard deviation due to viscous fingering, $\tilde{\sigma}_{a,f}$, as a function of time for $\kappa' = 0, 0.5, 3$.
From this figure, it is seen that the time, when $\tilde{\sigma}_{a,f}$ starts deviates from zero, varies non-monotonically with respect to $\kappa'$. Such a time was mentioned as the onset of VF by Mishra {\it et al.} \cite{Mishra2009}. In order to quantify this precisely, we define the onset time of fingers as,
\begin{equation}
t^c_{\rm f} = \min_{\tau}\left \{ \tau > t_0 : \tilde{\sigma}_{a,f}(\tau > t) = 0, ~ \tilde{\sigma}_{a,f}(\tau \leq t) > 0 \right\}. 
\end{equation}
In Fig. \ref{fig:onsetDNS}(a) solid dots ($\CIRCLE$) correspond to the onset time measured by visually observing fingers from the density plots of the solute concentration. For $R = 2, w = 512$ the dependence of $t^c_{\rm f}$ as a function of $\kappa'$ has been plotted in Fig. \ref{fig:onsetDNS}(b), which represents a non-monotonic dependence of $t^c_{\rm f}$ on $\kappa'$. It also shows the density plot of solute dynamics (see inset of Fig. \ref{fig:onsetDNS}(b)) at the onset of fingers, $t^c_{\rm f}$.  It can be shown that this non-monotonic characteristic of onset of instability can not be captured if $t^c$ is calculated from $\sigma_m$, instead of $\sigma_{mV}$. Thus, we conclude that the growth rate measured from the amplification measure of the solute concentration coupled with the velocity perturbation is more realistic than that corresponding to only solute concentration. Further, the effect of the log-mobility ratio $R$ is consistent with the classical VF instabilities in two component models \cite{Tan1986, Pramanik2013}, e.g., for a fixed $\kappa' \neq 0$ and wave number, instability sets in earlier for larger $R$ (see Fig. \ref{fig:onsetLSA}). 

\begin{figure}[h]
\centering
(a) \\
\includegraphics[width=3.5in, keepaspectratio=true, angle=0]{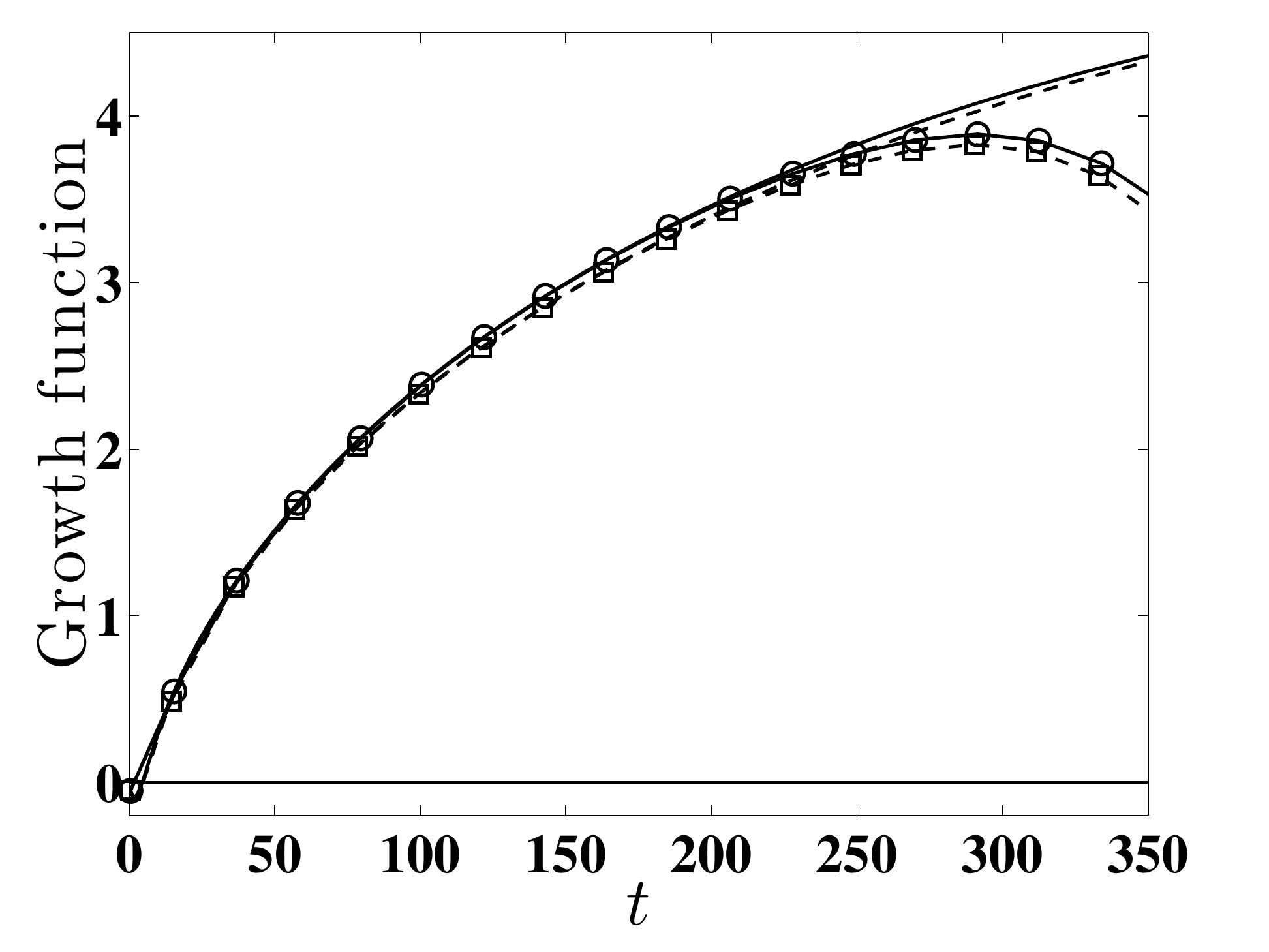} \\
(b) \\
\includegraphics[width=3.5in, keepaspectratio=true, angle=0]{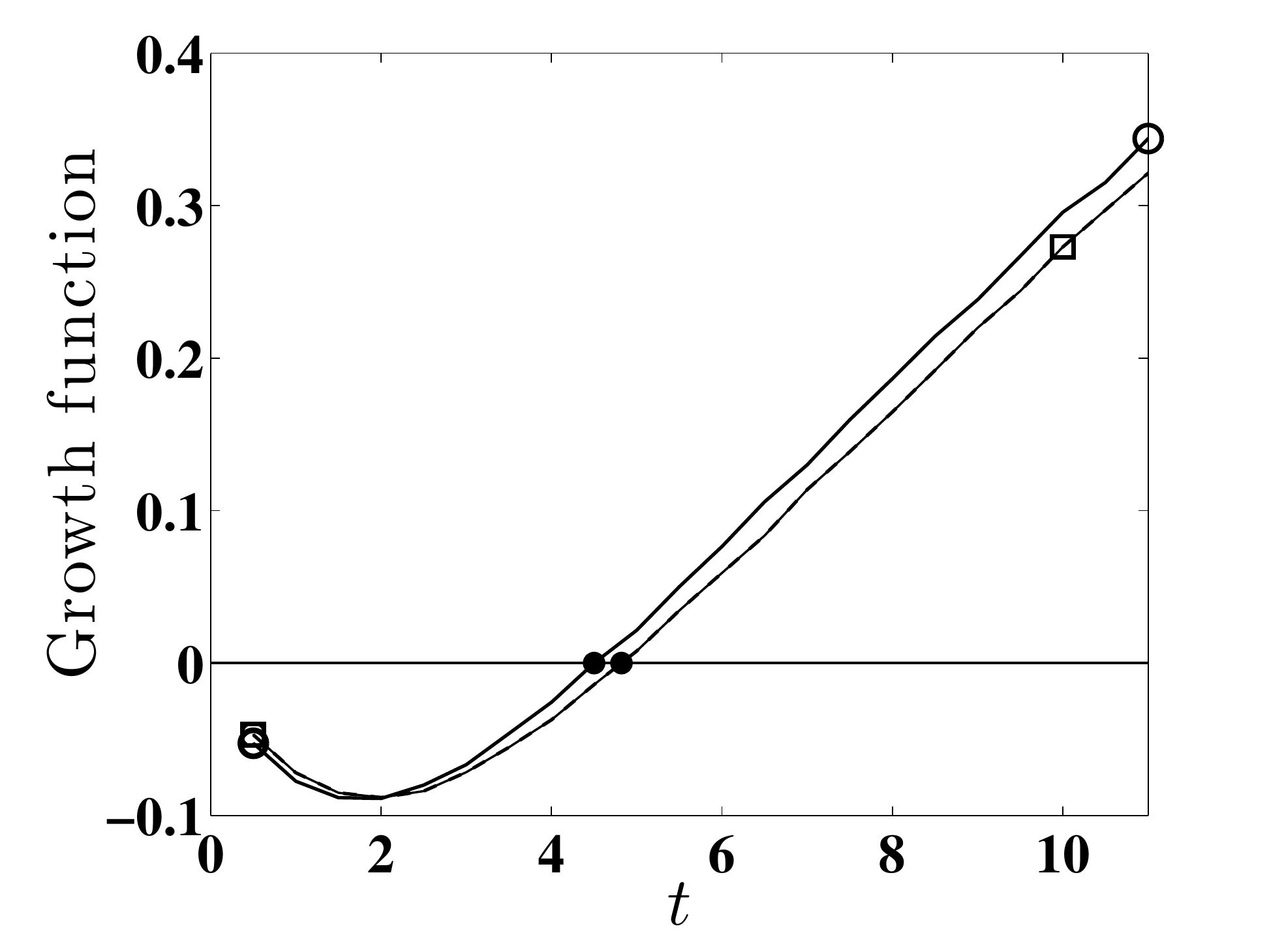}
\caption{(a) Growth functions, $\sigma_{mV} t$, of the solute concentration for $R = 2, w = 512, k = 0.1$ and $\kappa' = 0$ (solid line), $\kappa' = 0.2$ (dashed line): Lines without symbol correspond to LSA and lines with symbols represent DNS. (b) Magnified of (a) near the onset of  instability marked by the solid dots ($\CIRCLE$). }\label{fig:growth_kappa_0o2}
\end{figure} 

Mishra {\it et al.} \cite{Mishra2009} calculated that in absence of viscosity contrast between the displacing fluid and the sample solvent (i.e., $R$ = 0), the disengagement time of the solute from the solvent is given by (see Eq. (22) in \cite{Mishra2009})
\begin{equation}
t_{cric} = (1 + 1/\kappa')^2\bigg[\sqrt{2}(1 + \sqrt{\delta/(1 + \kappa')}) + \sqrt{2(1 + \sqrt{\delta/(1 + \kappa')^2}) + \kappa' w/(1 + \kappa')}\bigg]^2.
\end{equation}
 In the limiting case of very large $\kappa'$, $t_{cric}$ attains an asymptotic value $\left(\sqrt{2} + \sqrt{2 + w}\right)^2$. As an example, for a finite slice of width $w = 14$ this asymptotic value becomes $\approx 29$, which is very large compared to the onset of fingering instability $t^c$ measured from the present LSA. Thus the disengagement of the solute from the solvent zone is not possible before the linearly unstable modes set in for instability. However, our LSA captures the non-monotonicity in $t^c$. Thus we conclude that for all $\kappa' \neq 0 $ there exist linearly unstable modes, which may not develop into fingers in the nonlinear regime for large $\kappa'$. 

Next, we investigate the influence of $\kappa'$ both in the linear and nonlinear regimes and compare the results obtained with those in the absence of adsorption. In this context we choose $R = 2, w = 512, k = 0.1$ and $\kappa' = 0, 0.2$. The temporal evolution of the growth functions obtained from both LSA and DNS are shown in Fig. \ref{fig:growth_kappa_0o2}. Fig. \ref{fig:growth_kappa_0o2}(b) illustrates that when diffusion dominates at the early time, the growth functions $\sigma_{mV} t$ corresponding to two values of $\kappa'$ are almost identical (see curves for $ t \leq 2$). As soon as they start growing in the convection dominated regime they are different (see curves for $ t > 2$), and this leads to different onset time. However, it is identified that the onset of non-linearity is almost independent of the retention parameter $\kappa'$ (see Fig. \ref{fig:growth_kappa_0o2}(a)). 

\section{Conclusion}\label{sec:conclusion}
We have theoretically studied the onset of fingering instability in a finite slice of linearly adsorbed solute. Instability is driven by the viscosity contrast between the displacing fluid and the sample solvent containing the solute. We presented a linear stability analysis based on a Fourier pseudo-spectral method, which, compared to QSSA methods, captures both the onset of instability and the early time diffusion dominated regime. The linearized equations are solved as an initial value problem and the growth rate associated with each perturbation quantities is calculated from their respective amplification measure. The numerical results revealed that the exponential growth of the perturbations are reasonable in the linear regime. It is verified that, in absence of retention, i.e., $\kappa' = 0$, the onset time is independent of $w \geq w_c$, a critical value. Further, it is shown that there exists a threshold finite slice width, beyond which the stability analysis for finite sample is identical to that of a single interface displacement. Another very interesting observation is that, the onset time is a non-monotonic function of the retention parameter, $\kappa'$. It is shown that for a given wave number the largest onset time and the associated retention parameter decrease as log-mobility ratio increases. The present LSA agrees qualitatively with DNS and we successfully distinguish between the linear and nonlinear regimes. Analysis with velocity dependent dispersion and nonlinear adsorption isotherm has been undertaken for further study. 

\section*{Acknowledgements}
S.P. acknowledges the National Board for Higher Mathematics, Department of Atomic Energy, Government of India for the Ph.D. fellowship. 

\appendix

\section{Algorithm of the present IVP approach for LSA and DNS}\label{sec:Algo}
The present linear stability analysis is of generic type as it handles the unsteady base-state very carefully that helps to capture the underlying physics more appropriately. Below we describe the algorithm of the numerical method used in the present LSA: 
\begin{enumerate}
\item We introduce perturbation to $c, c_{a,m}$ and $\psi$ at time $t = t_0$ and compute the coefficients in Eqs. \eqref{eq:LP1}-\eqref{eq:LP3} at $t = t_0$. 
\item Time integration is performed by taking the perturbations as the initial condition to the unknown variables, $c', c'_{a,m}$ and $\psi'$. 
\item Obtained solutions, $c', c'_{a,m}$ and $\psi'$ are used as the initial condition for the time marching in the next step. Repeat this step until desired result is obtained. 
\end{enumerate}

\section{}
Most of the LSA methods existing in the literature of fingering instabilities driven by viscosity contrast solve the following linearized equations,
\begin{eqnarray}
\label{eq:A1}
& & \left[ \frac{\partial^2}{\partial x^2} + \frac{\partial^2}{\partial y^2} + R\frac{\partial c^b}{\partial x}\frac{\partial}{\partial x}\right]u' = -R\frac{\partial^2 c'}{\partial y^2}, \\
\label{eq:A2}
& & \left[\frac{\partial } {\partial t} - \frac{\partial^2}{\partial x^2}  - \frac{\partial^2}{\partial y^2}  \right] c' = -\frac{\partial c^b}{\partial x} u', \\
\label{eq:A3}
& &  \left[\frac{\partial } {\partial t} - \delta\frac{\partial^2}{\partial x^2}  - \delta\frac{\partial^2}{\partial y^2} + \lambda \frac{\partial }{\partial x} \right] c_{a,m}' = -\delta\frac{\partial c_{a,m}^b}{\partial x} u'. 
\end{eqnarray}
Since, the coefficients of Eqs. \eqref{eq:A1}-\eqref{eq:A3} are independent of $y$, wave like disturbances are assumed of the form, 
\begin{equation}
\label{eq:A4}
(u', c', c_{a,m}') (x, y,t) = (\Psi, \Phi, \Theta) (x,t) \exp(i k y ),
\end{equation}
where $k$ represents the non-dimensional wave number in $y$-direction. The resultant linear equations can be written compactly as a non-autonomous IVP
\begin{equation}
\label{eq:A5}
\frac{\partial \textbf{q}(t)}{\partial t} =  A(t) \textbf{q}(t), ~~~ \textbf{q}(x,t_0) = \textbf{q}_0(x),
\end{equation} 
where
\begin{eqnarray}
& & \textbf{q} = \begin{bmatrix}
\Phi\\
\Theta
\end{bmatrix}, ~~~
A(t) = \begin{bmatrix}
M_3 - M_4M_1^{-1}M_2 & 0\\
-M_6M_1^{-1}M_2 & M_5
\end{bmatrix}, \nonumber \\
& & M_1 = \mathcal{D}  + R\frac{\partial c^b}{\partial x} \frac{\partial}{\partial x}, ~ M_2 = Rk^2\mathbf{I}, ~ M_3 = \frac{\partial^2}{\partial x^2} - k^2\mathbf{I} (\equiv \mathcal{D}), \nonumber \\
& & M_4 = -\frac{\partial c^b}{\partial x}, ~ M_5 = \delta\mathcal{D} - \lambda \frac{\partial c'_{a,m}}{\partial x}, ~ M_6 = -\delta  \frac{\partial c_{a,m}^b}{\partial x}, \nonumber
\end{eqnarray}
$\mathbf{I}$ is the identity operator, and $t_0$ is the time at which the perturbations are introduced.

\subsection{Quasi-steady state approximation method}\label{subset:QSSA}
This classical frozen time approach of investigating the instability assumes that the unsteady base state evolves very slowly in comparison to the perturbations. Rewriting Eq. \eqref{eq:A4} as $(u', c', c_{a,m}') (x, y,t) =  (\Psi, \Phi, \Theta) (x) \exp(i k y  + \sigma(t_q) t)$, where $t_q$ is the time at which the unsteady base states (see Eqs. \eqref{concen_base_state} and \eqref{solute_base_state}) are frozen, Eq. \eqref{eq:A5} reduces to an algebraic eigenvalue problem, i.e., $A(t) \textbf{q}(x) = \sigma(t_q)\textbf{q}(x) $. The maximum eigenvalue of $(A + A^t)/2$ can be interpreted as the maximum possible instantaneous growth rate \cite{Trefethen2005} that can be achieved by any initial condition at early times. This reveals that QSSA does not capture separate growth rates for the solvent and solute concentration perturbations, thus restricting the analysis of present model to that of classical VF instabilities \cite{Pramanik2013, Kim2012}. However, the DNS results of Mishra {\it et al.} \cite{Mishra2009} showed that the retention parameter $\kappa'$ influences the instability of the solute, but not the solvent. Therefore, the underlying instability dynamics of solute can not be captured by the QSSA method. Hence, an IVP approach is imperative to find the LSA of adsorbed solute transport. 

\subsection{IVP approaches}\label{subsec:IVP}
Due to unsteady nature of the base state flow the growth rate obtained by solving the IVP, Eq. \eqref{eq:A5}, is sensitive to the initial condition $\textbf{q}(x,t_0) = \textbf{q}_0(x)$. In the literature, the IVPs are solved using an initial condition that corresponds to a random perturbation in the whole spatial domain,
\begin{eqnarray}
\label{eq:A6}
\textbf{q} (x,t_0)= \epsilon*\text{rand}(x), ~~~~ \forall x,
\end{eqnarray}
where $\text{rand}(x)$ represents a random number generator between $-1$ and $1$, and $\epsilon$ corresponds to the amplitude of the perturbation, which is a very small positive number ($0 < \epsilon \ll 1$). Since it is known that the fastest growing perturbation is localized around the diffusive interface \cite{Ben2002}, we solve Eq. \eqref{eq:A5} in $(x,y,t)$ coordinate system with an initial condition of the form, 
\begin{eqnarray}
\label{eq:A7}
\textbf{q} (x,t_0) =
\begin{cases}
\epsilon*\text{rand}(x), &  a_1 \leq x \leq b_1 \\
0, & ~ \mbox{otherwise}.
\end{cases}
\end{eqnarray}
Here the spatial interval $[a_1, b_1]$ corresponds to the thickness of the diffusive layer, and its location depends on the viscosity contrast between the fluids. The results obtained have very good agreement with the DNS results and the present pseudo-spectral method based LSA. 


\end{document}